\documentclass[aps,prb,reprint,superscriptaddress,longbibliography,floatfix]{revtex4-2}

\usepackage[utf8]{inputenc}
\usepackage[english]{babel}
\usepackage{graphicx}
\usepackage{amsmath,amssymb}
\usepackage{booktabs}
\usepackage{microtype}
\usepackage{bm}
\usepackage[colorlinks=true,linkcolor=blue,citecolor=blue,urlcolor=blue]{hyperref}
\usepackage{newunicodechar}

\DeclareUnicodeCharacter{2212}{-}
\newunicodechar{∣}{\mid}
\begin{document}

\title{Local-dynamics accessibility of the stripe phase in the triangular-lattice $J_1$--$J_2$ Ising antiferromagnet}

\author{Tom Poisblaud}
\affiliation{Independent Researcher}

\begin{abstract}
The interplay between energetic relaxation and global stripe-sector selection is investigated by analyzing the finite-rate cooling dynamics of the triangular-lattice $J_1\text{–}J_2$ Ising antiferromagnet under local single-spin Metropolis updates. A protocol-dependent stripe-accessibility time, $n^*(L, J_2/J_1)$, was established as an operational diagnostic to determine the threshold at which half of the trajectories achieve global stripe order. Real-space morphology diagnostics and fixed-temperature holds were subsequently employed to assess late-stage domain wall elimination and to characterize the finite-time coarsening behavior across varying system sizes and $J_2/J_1$ coupling ratios. The system quickly forms local stripe patterns before aligning them globally. Final alignment is slowed by domain-wall elimination, taking longer in larger systems but speeding up with a higher $J_2/J_1$ ratio.
\end{abstract}

\maketitle

\section{Introduction}

A thermodynamically stable phase is not necessarily dynamically accessible on finite time scales. This distinction is particularly important in geometrically frustrated systems~\cite{Balents2010,Lacroix2011}, where local constraints can restrict the pathways connecting competing configurations. The triangular-lattice Ising antiferromagnet provides a paradigmatic setting in which to investigate this distinction. Because the triangular lattice is not bipartite, its antiferromagnetic nearest-neighbor interactions cannot all be satisfied simultaneously, leading to an extensively degenerate ground-state manifold with finite residual entropy and algebraic correlations at zero temperature~\cite{Wannier1950,Stephenson1970,BloteHilhorst1982,BloteNightingale1993,Mila2015}.

The nearest-neighbor ground-state manifold can be represented in terms of dimers, strings, or height variables, and local dynamics within this manifold is constrained by a sector structure~\cite{BloteHilhorst1982,BloteNightingale1993,Kim2007}. Consequently, local cooling can reduce the energy and restore the local antiferromagnetic constraint without necessarily producing the global rearrangements required to select a unique stripe sector.

Adding an antiferromagnetic next-nearest-neighbor interaction \(J_2\) lifts the extensive degeneracy and favors a sixfold-degenerate collinear stripe phase at low temperature. The equilibrium stripe phase and its first-order transition from the paramagnetic phase are well established~\cite{Rastelli2005,Korshunov2005,Smerald2016}. The question addressed here is therefore not whether the stripe phase is thermodynamically favored, but whether local dynamics can reach it during a finite-rate cooling protocol. More precisely, we ask: how does a system evolve toward the stripe phase selected by \(J_2/J_1>0\), and which part of this ordering process becomes inaccessible when the cooling is too fast or \(J_2/J_1\) is too small?

This question is particularly relevant in the weak-coupling regime \(J_2/J_1\to0^+\), where the energetic driving force toward stripe order becomes weak while the system remains influenced by the frustrated nearest-neighbor manifold.

Because the equilibrium transition into the stripe phase is first order, finite-rate ordering is not expected to be described by the critical slowing down of a conventional continuous transition. Instead, metastability and the motion and coarsening of domain walls between competing stripe orientations can become important~\cite{Binder1987,Bray1994}. We therefore focus on the finite-time accessibility of the known equilibrium stripe phase under fixed local Metropolis dynamics, rather than on its equilibrium phase diagram or a complete nucleation theory.

We investigate this problem using fixed local single-spin Metropolis dynamics and finite-rate cooling. We characterize the resulting finite-time accessibility through the operational boundary \(n^*(L,J_2/J_1)\), and combine it with energy, local-constraint, and real-space diagnostics to identify the stage of ordering that limits global stripe selection.

We find that finite-time failure is not simply associated with incomplete energetic relaxation: trajectories can reach low-energy, locally stripe-organized configurations while retaining domain walls and competing stripe orientations. Global stripe selection, therefore, constitutes a distinct, slower dynamical stage of the ordering process.

Finite-rate ordering is often discussed in the framework of the Kibble--Zurek mechanism~\cite{Kibble1976,Zurek1985,Zurek1996, Dziarmaga2010,DelCampoZurek2014}. Since the transition considered here is first order, however, the standard Kibble--Zurek scaling arguments do not apply directly~\cite{Rastelli2005,Korshunov2005}.

The model is defined by the Hamiltonian
\begin{equation}
H = J_1 \sum_{\langle i,j\rangle} s_i s_j + J_2 \sum_{\langle\langle i,j\rangle\rangle} s_i s_j,
\label{eq:hamiltonian}
\end{equation}
with $J_1>0$, $J_2>0$, and $s_i=\pm1$. The sums run over nearest-neighbor and next-nearest-neighbor pairs on the triangular lattice with periodic boundary conditions. We use \(J_1=1\) as the energy unit.

\section{Methods}\label{sec:methods}

The system described by Eq.~\eqref{eq:hamiltonian} is simulated on an \(L\times L\) triangular lattice with periodic boundary conditions, using \(N=L^2\) spins. The dynamics consist of local single-spin Metropolis updates~\cite{Metropolis1953}. One Monte Carlo sweep corresponds to one attempted flip per spin. A proposed flip is accepted with probability \(\min[1,\exp(-\Delta E/T)]\), where \(\Delta E\) is the energy change associated with the flip. The update schedule uses a fixed colored-sublattice implementation of the same local Metropolis rule; the lattice offsets, color classes, and validation checks are given in Appendix~\ref{app:numerical_details}.

The main cooling protocol starts from a random high-temperature configuration at $T_{\mathrm{init}}=3.0$ and decreases the temperature down to $T_{\mathrm{final}}=0.05$ with a fixed step $dT=0.05$. At each temperature, the system is evolved for $n_{\mathrm{sweeps}}/T$ sweeps, and the final configuration at one temperature is used as the initial condition at the next lower temperature. The global finite-rate datasets use $L=64,96,128,160,200$, $J_2/J_1=0.08,0.09,0.10,0.11,0.15$, and $n_{\mathrm{sweeps}}/T=500$ to $20000$. Most points contain $N_{\mathrm{seeds}}=50$ independent trajectories, while the boundary regions at $J_2/J_1=0.09$ and $0.10$ were reinforced up to $N_{\mathrm{seeds}}=125$ for the slowest rates.

the dominant stripe order is computed from the structure factor at the three symmetry-related $M$ points. In the reciprocal coordinates conjugate to the integer lattice coordinates, these points can be taken as $\mathbf M_1=(\pi,0)$, $\mathbf M_2=(0,\pi)$, and $\mathbf M_3=(\pi,\pi)$. The structure factor is normalized as
\begin{equation}
S(\mathbf q)=\frac{1}{N^2}\left|\sum_j s_j e^{i\mathbf q\cdot\mathbf r_j}\right|^2.
\end{equation}
the dominant stripe order is then defined by
\begin{equation}
M_{\mathrm{stripe}}=\max_{a=1,2,3} S(\mathbf M_a).
\end{equation}
With this convention, $M_{\mathrm{stripe}}\simeq1$ denotes a perfect single-orientation stripe state and remains small in a disordered state. The final order parameter is denoted by $M_f=M_{\mathrm{stripe}}(T_{\mathrm{final}})$.

The central probability used to define the kinetic boundary is
\begin{equation}
P_{\mathrm{stripe}}=P(M_f\geq0.5).
\end{equation}
The operational stripe-accessibility time $n^*(L,J_2/J_1)$ is defined by the condition $P(M_f\geq0.5)=0.5$. Since the measured probabilities are noisy and should increase monotonically with $n_{\mathrm{sweeps}}/T$ at fixed $L$ and $J_2/J_1$, the boundary is extracted by applying isotonic regression to $P_{\mathrm{stripe}}(n_{\mathrm{sweeps}}/T)$ and then interpolating the value at which the fitted curve crosses $0.5$. Uncertainties on $n^*$ are estimated by bootstrap resampling of independent seeds: for each bootstrap sample, the full extraction procedure, including isotonic regression and interpolation, is repeated. Unless an explicit command-line value was used, the analysis uses $N_{\mathrm{boot}}=100$ bootstrap resamples; production runs for which the command-line option was specified used $N_{\mathrm{boot}}=200$. The values reported in the tables are central estimates. Bootstrap intervals are shown as error bars in the corresponding boundary figures whenever the interval is resolved; censored estimates are reported as inequalities. The resulting $n^*(L,J_2/J_1)$ should be interpreted as a kinetic accessibility threshold for the chosen protocol and update rule.

\begin{table*}[t]
\centering
\small
\begin{tabular}{c c c}
\toprule
Symbol & Criterion & Use \\
\midrule
\(P_{\mathrm{stripe}}\) & \(M_f\geq0.5\) & Main kinetic boundary \(n^*(L,J_2/J_1)\) \\
\(P_{\mathrm{stripe}}^{\mathrm{strict}}\) & \(M_{\max}>0.7,\ e-e_{\mathrm{stripe}}<0.03\) & Kinetic trapping and hold diagnostics \\
\(P_{\mathrm{stripe}}^{E,M}\) & \(M_f>0.75,\ e_f-e_{\mathrm{stripe}}<0.1\,J_2/J_1\) & Weak-\(J_2\) scan \\
\(P_{\mathrm{single}}\) & \(M_{\max}\geq0.7,\ \rho_{\mathrm{DW}}\leq0.15\) & Real-space single-domain morphology \\
\bottomrule
\end{tabular}
\caption{Operational stripe-success criteria used in this work. The criteria are deliberately protocol-dependent and are used for different purposes. \(P_{\mathrm{stripe}}\) defines the main kinetic boundary, \(P_{\mathrm{stripe}}^{\mathrm{strict}}\) is used in the trapping diagnostics, and \(P_{\mathrm{stripe}}^{E,M}\) is used in the weak-\(J_2\) perturbative analysis.}
\label{tab:success_criteria}
\end{table*}

Because the different sections address different physical questions, several operational success criteria are used. They are summarized in Table~\ref{tab:success_criteria}. The main kinetic boundary uses the least restrictive global structure-factor criterion, while the mechanism, morphology, and weak-$J_2/J_1$ analyzes use stricter criteria adapted to their specific diagnostic purposes.

Additional diagnostics are used to interpret the final configurations. The orientation-resolved stripe intensities are denoted $M_1$, $M_2$, $M_3$, and $M_{\max}=\max(M_1,M_2,M_3)$. The orientation-competition parameter is
\begin{equation}
C_{\mathrm{orient}}=\frac{M_1+M_2+M_3-M_{\max}}{M_1+M_2+M_3},
\end{equation}
which is smaller for a single dominant stripe orientation and larger when several orientations coexist. A domain-wall-density proxy \(\rho_{\mathrm{DW}}\) is measured in the gage of the dominant stripe orientation, together with the stripe-sector proxy \(W_{\max}^{\mathrm{proxy}}=LM_{\max}\) and the density \(\rho_{\triangle}^{\mathrm{defect}}\) of elementary triangles with three equal spins. When $\rho_{\triangle}^{\mathrm{defect}}\ll1$, frustrated nearest-neighbor bonds are mapped to dimers on the dual honeycomb lattice and used to compute a dimer-winding diagnostic.

For the mechanism analysis, additional runs were performed at $L=128$ and $J_2/J_1=0.10$ with $n_{\mathrm{sweeps}}/T=1500$ to $14000$ and $N_{\mathrm{seeds}}=40$. These runs stored final configurations and measured a stripe-domain-length proxy $\xi_{\mathrm{stripe}}$. Fixed-temperature holds, size-dependence tests, and simple collective-move comparisons were used as diagnostic protocols to distinguish thermodynamic instability from finite-time kinetic trapping. These auxiliary protocols are not intended to replace equilibrium simulations; they are used to identify which relaxation processes are limiting the local-dynamics cooling trajectories.

\section{Finite-rate accessibility of the stripe phase}\label{subsec:finite_rate_stripe_selection}

We first determine how finite-rate cooling controls access to the globally ordered stripe phase. We focus on the dependence of the final stripe order on cooling time, system size, and $J_2/J_1$, and use this dependence to identify the characteristic time scale for global stripe selection. The cooling protocol is controlled by the number of Monte Carlo sweeps performed at each temperature step, $n_{\mathrm{sweeps}}/T$, with larger values corresponding to slower cooling. The final stripe order is defined as
\begin{equation}
M_f \equiv M_{\mathrm{stripe}}(T_{\mathrm{final}}),
\end{equation}
where $M_{\mathrm{stripe}}$ is obtained from the largest structure-factor intensity at the three $M$-point orientations and is normalized so that $M_{\mathrm{stripe}}\simeq1$ for a perfect stripe state. In addition to the mean value of \(M_f\), the finite-rate stripe-accessibility probability is defined as
\begin{equation}
P_{\mathrm{stripe}} =
P(M_f\geq0.5),
\label{eq:pstripe_definition}
\end{equation}
The kinetic accessibility scale \(n^*(L,J_2/J_1)\) is defined as the value of $n_{\mathrm{sweeps}}/T$ at which
\begin{equation}
P_{\mathrm{stripe}}\!\left(n^*\right)=0.5.
\label{eq:nstar_definition}
\end{equation}

The corresponding operational stripe-accessibility time, expressed in sweeps per temperature value step, is denoted $n^*(L,J_2/J_1)$. This definition is intentionally kinetic: it describes the dynamical accessibility of the stripe state under the specified cooling protocol and local update rule.

Although \(J_2/J_1>0\) thermodynamically selects the stripe phase, local Metropolis dynamics need not establish a single global stripe orientation within the available cooling time. A final configuration can therefore contain locally stripe-ordered domains while retaining a small global \(M_f\) because competing orientations and domain walls have not yet been eliminated.

We performed finite-rate cooling simulations for $L=64-256$, $J_2/J_1=0.080.15$ and $n_\mathrm{sweeps}=500-60000$ per temperature steps. The data sets contain 50 independent trajectories with additional sampling near the accessibility boundary. The accessibility boundary is extracted from \(P_{\mathrm{stripe}}\) using isotonic regression to impose the physically motivated monotonic increase of $P_\mathrm{stripe}$ with cooling time \(n_{\mathrm{sweeps}}/T\). Uncertainties are estimated by bootstrap resampling.

\begin{figure*}[t]
\centering
\includegraphics[width=\textwidth]{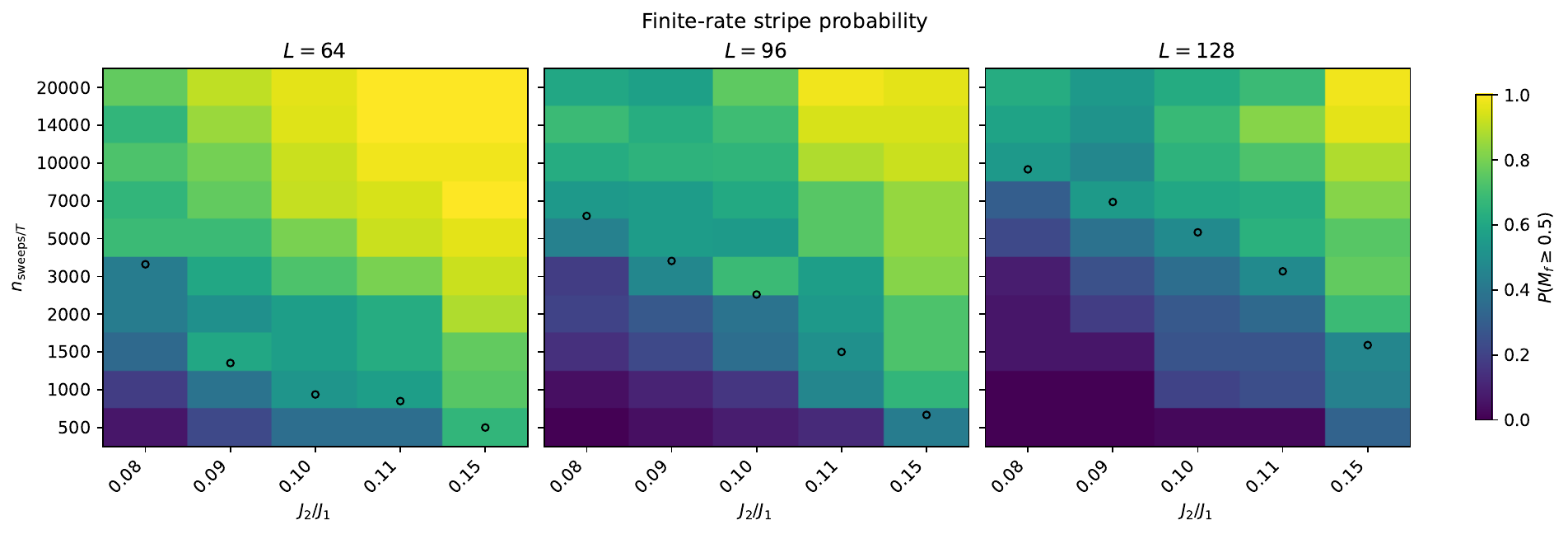}
\caption{Finite-rate stripe probability $P(M_f\geq0.5)$ in the $(J_2/J_1,n_{\mathrm{sweeps}}/T)$ plane for $L=64,96,128$. Markers indicate the extracted kinetic boundary $n^*(L,J_2/J_1)$ obtained after isotonic regression. Increasing $n_{\mathrm{sweeps}}/T$ or $J_2/J_1$ promotes global stripe selection, while the extracted boundary shifts toward longer cooling time as the system size increases.}
\label{fig:axe1_heatmaps}
\end{figure*}

Figure~\ref{fig:axe1_heatmaps} shows $P_{\mathrm{stripe}}$ in the $(J_2/J_1,n_{\mathrm{sweeps}}/T)$ plane for the main sizes. The resulting boundary shows two robust trends. First, slower cooling increases the probability of reaching a globally stripe-ordered state. Second, increasing \(J_2/J_1\) shifts the boundary toward faster cooling, showing that a stronger energetic bias facilitates global stripe selection. The same behavior persists at larger system sizes, with the accessibility boundary shifting toward slower cooling as $L$ increases..

The extracted thresholds are summarized in Table~\ref{tab:axe1_boundary_extended}. At fixed \(L\), \(n^*\) decreases as \(J_2/J_1\) increases. At fixed \(J_2/J_1\), the refined data now show a consistent increase of \(n^*\) with system size, confirming that global stripe selection remains costly at large sizes.

The measured \(n^*\) increases strongly with system size. At \(J_2/J_1=0.10\), the available data are consistent, over the simulated size range, with a growth of $n^*$ of order \(L^2\), but the present data do not establish an asymptotic dynamic exponent. Weaker couplings, however, show stronger deviations and broader crossover behavior. These results indicate that global stripe selection becomes progressively more time-consuming as the system size increases. An empirical finite-window parameterization of the boundary and its fitted parameters are reported in Appendix~\ref{sub;effective_fit}.

\begin{figure}[t]
\centering
\includegraphics[width=0.5\textwidth]{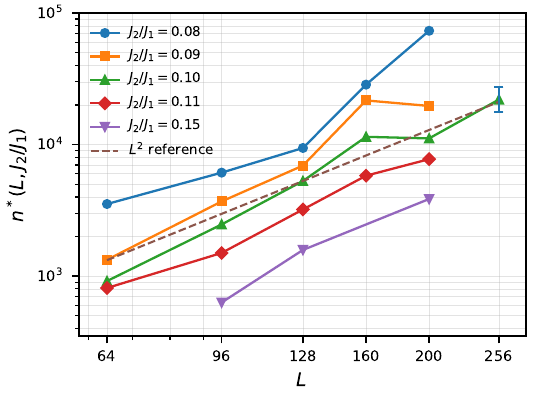}
\caption{Size dependence of \(n^*(L,J_2/J_1)\) after inclusion of the \(L=160\), \(L=200\), and targeted \(L=256\) data. The \(L=256\) point at \(J_2/J_1=0.10\) lies close to the quadratic extrapolation trend suggested from \(L\leq128\), while the full finite-size sequence remains affected by broad probabilistic crossings and finite-window corrections.}
\label{fig:axe1_size_scaling}
\end{figure}

The resulting accessibility boundary has a clear kinetic character: increasing the available cooling time promotes global stripe selection, whereas increasing system size makes complete ordering progressively more difficult. Increasing $J_2/J_1$ shifts the boundary toward faster cooling, indicating that a stronger energetic bias facilitates global stripe selection. The size dependence suggests that the limiting process involves a collective ordering timescale rather than purely local stripe formation. The real-space analysis below tests this interpretation by directly tracking domain growth and domain-wall removal.

\begin{figure}[t]
\centering
\includegraphics[width=0.5\textwidth]{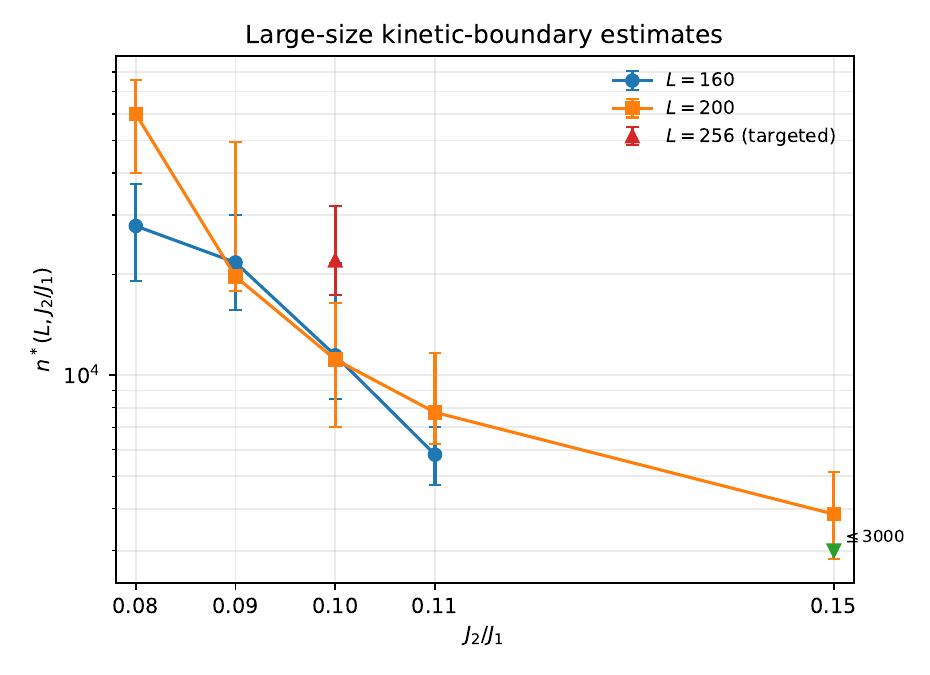}
\caption{Large-size kinetic-boundary estimates. The \(L=160\) and \(L=200\) data cover several couplings, while the \(L=256\) point is a targeted check at \(J_2/J_1=0.10\). Error bars show bootstrap intervals obtained from 200 resamples. The weakest \(L=200\) coupling, \(J_2/J_1=0.08\), crosses only near the upper end of the simulated window, and the \(L=256\), \(J_2/J_1=0.10\) point confirms that global stripe selection remains costly at large size. The \(L=160\), \(J_2/J_1=0.15\) point is left-censored, with \(n^*(160,0.15)\leq 3000\).}
\label{fig:axe1_large_size_boundaries}
\end{figure}

\section{From local stripe formation to global stripe selection}~\label{sec:local_stripe_selection}

\subsection{Diagnostic protocols}

The previous section established an operational stripe-accessibility time, \(n^*(L,J_2/J_1)\), from the final stripe probability.
We now determine whether the remaining disorder is associated with domain-wall mobility, competing stripe orientations, or incomplete relaxation within the frustrated manifold.

The analysis uses the orientation-resolved stripe intensities \(M_1,M_2,M_3\), $M_{\max}$ and the orientation-competition parameter $C_{\rm orient}$. Large \(M_{\max}\), small \(C_{\rm orient}\), and small \(\rho_{\rm DW}\) indicate a clean single-orientation stripe state, whereas large \(C_{\rm orient}\) or \(\rho_{\rm DW}\) indicate competing orientations or residual walls. The stripe-sector proxy \(W_{\max}^{\rm proxy}=LM_{\max}\), the triangular-defect density \(\rho_{\triangle}^{\rm defect}\), and, when \(\rho_{\triangle}^{\rm defect}\ll1\), the dimer-winding diagnostic \((W_1^{\rm dimer},W_2^{\rm dimer})\) are used.

For the stricter near-ground-state criterion used in this section, a configuration is counted as successful when
\begin{equation}
M_{\max}>0.7,\qquad e-e_{\rm stripe}<0.03,
\end{equation}
with
\begin{equation}
e_{\rm stripe}=-1-\frac{J_2}{J_1}.
\end{equation}
This criterion is stricter than \(P_{\mathrm{stripe}}=P(M_f\geq0.5)\) and is used only to diagnose kinetic trapping near the stripe ground-state manifold.

The auxiliary mechanism protocol cools from \(T_{\max}=1.5\) to \(T_{\min}=0.1\) using the same local Metropolis dynamics and the same definition of \(n_{\mathrm{sweeps}}/T\). This reduced temperature window is not used to define the global boundary \(n^*(L,J_2/J_1)\); it is a diagnostic protocol designed to focus on the low-temperature mobility regime where local stripe correlations, domain-wall motion, and orientation selection compete.

\subsection{Energetic relaxation and emergence of local stripe order}\label{subsec:energy_relaxation_and_local_stripe_formation}

\begin{figure*}[t]
\centering
\includegraphics[width=0.32\textwidth]{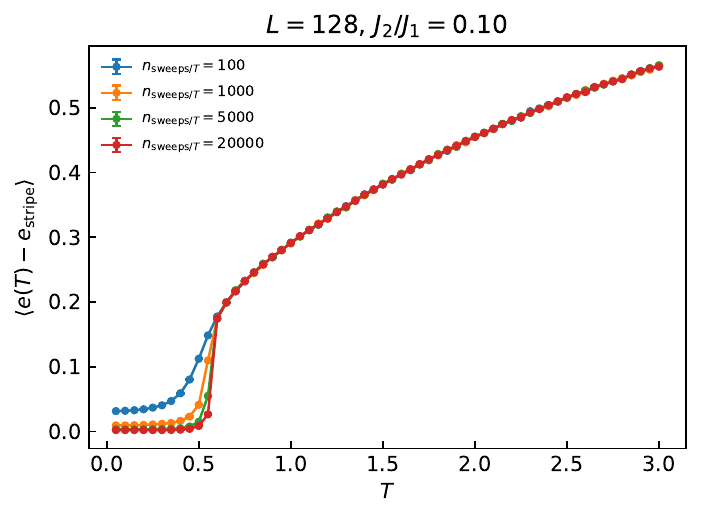}
\includegraphics[width=0.32\textwidth]{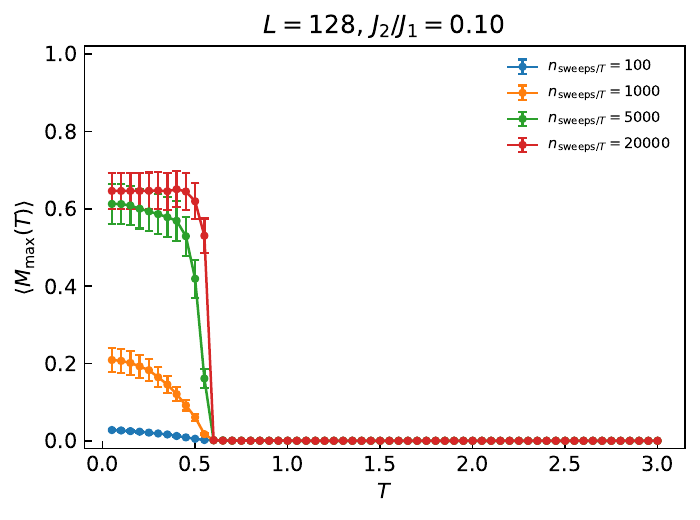}
\includegraphics[width=0.32\textwidth]{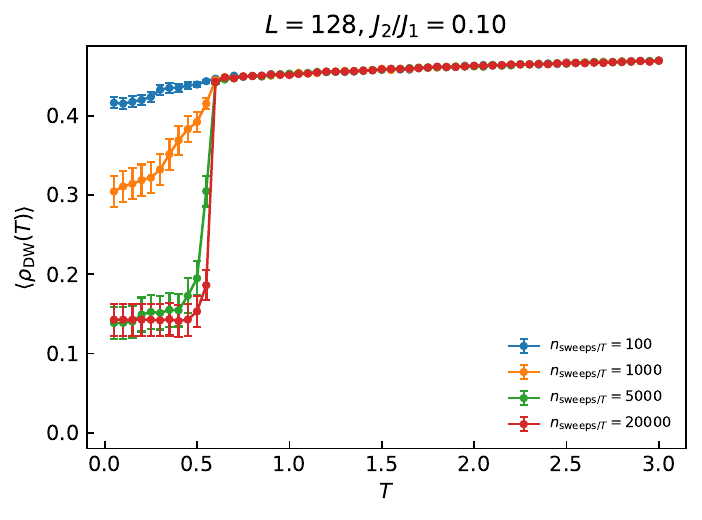}
\caption{Temperature-resolved energetic and ordering response during finite-rate cooling at \(L=128\) and \(J_2/J_1=0.10\). Left: excess energy relative to the stripe ground-state energy, \(\langle e(T)-e_{\mathrm{stripe}}\rangle\). Middle: dominant stripe order \(\langle M_{\max}(T)\rangle\). Right: dominant-gage domain-wall density \(\langle\rho_{\mathrm{DW}}(T)\rangle\). Different curves correspond to different values of \(n_{\mathrm{sweeps}}/T\), and error bars are standard errors over \(50\) independent seeds. Slower cooling lowers the final excess energy, increases global stripe order, and suppresses residual domain walls.}
\label{fig:energy_order_wall_response}
\end{figure*}

To separate energetic relaxation from global stripe selection, we analyze the cooling trajectories as a function of temperature. Figure~\ref{fig:energy_order_wall_response} shows the excess energy relative to the stripe ground-state energy, the dominant stripe order, and the dominant-gage domain-wall density for \(L=128\), \(J_2/J_1=0.10\), and several cooling rates. The energy reference is
\begin{equation}
e_{\mathrm{stripe}}=-1-\frac{J_2}{J_1}.
\end{equation}
For the present normalization of the Hamiltonian, this is the energy per spin of a perfect stripe configuration.

The energy decreases over the same temperature window in which stripe order grows, but the two diagnostics do not carry identical information. Faster cooling leaves larger excess energy, smaller \(M_{\max}\), and higher residual domain-wall density. Thus, trajectories that fail to reach global stripe order are not simply high-energy disordered states:
they contain locally stripe-ordered domains whose coarsening and orientation selection remain incomplete.

\subsection{Domain formation and late-stage coarsening}

A run was performed at $L=128$ and $J_2/J_1=0.10$, using $n_{\mathrm{sweeps}}/T=1500$ to $14000$ and $N_{\mathrm{seeds}}=40$. This run kept final configurations and measured a stripe-domain-length proxy $\xi_{\mathrm{stripe}}$. The extracted midpoint is $n^*(128,0.10)\simeq5000$, consistent with the value $n^*(128,0.10)\simeq5288$ obtained from a larger global dataset. The summary in Table~\ref{tab:axe1_mechanism} shows that the distribution of outcomes near the boundary is broad: at $n_{\mathrm{sweeps}}/T=5000$, $P(M_f\geq0.5)=0.50$, while the standard deviation of $M_f$ remains about $0.33$.

\begin{figure*}[t]
\centering
\includegraphics[width=\textwidth]{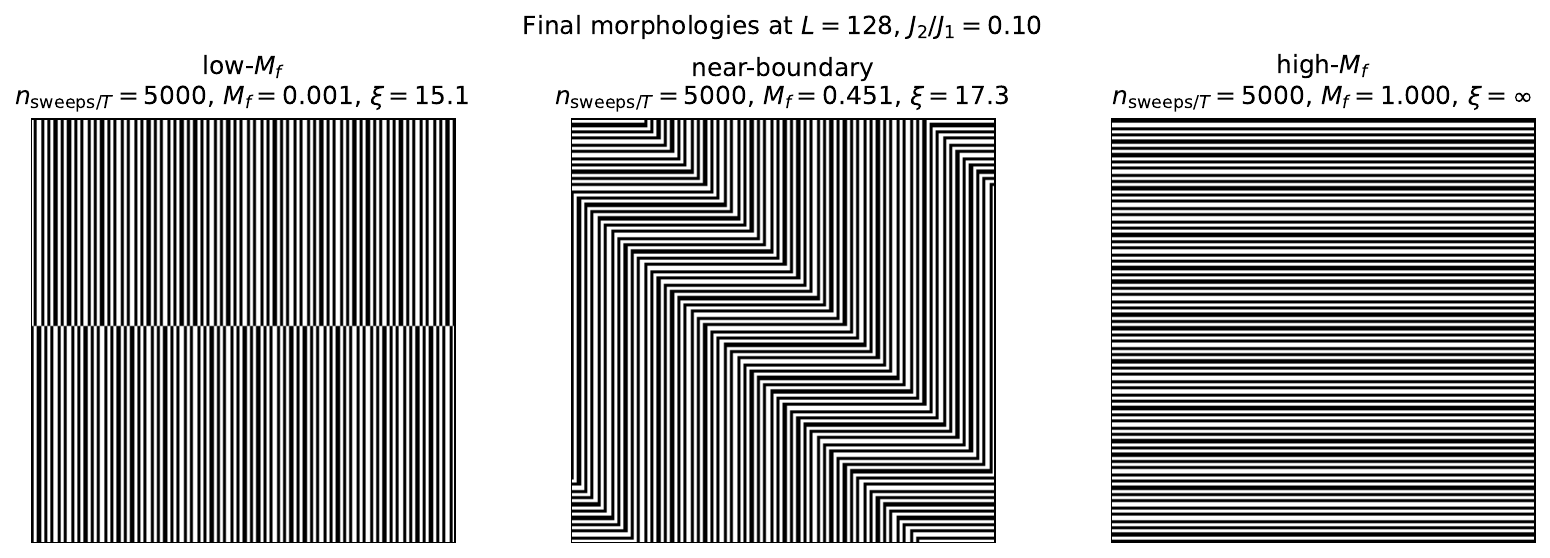}
\caption{Representative final configurations at $L=128$, $J_2/J_1=0.10$, and $n_{\mathrm{sweeps}}/T=5000$, close to the kinetic boundary. Independent trajectories at the same parameters can end in low-$M_f$, near-boundary multidomain, or saturated stripe states. The panels show that local stripe correlations may be present even when the global structure-factor order remains small because domains, phases, or orientations compete.}
\label{fig:axe1_morphology}
\end{figure*}

The final configurations in Fig.~\ref{fig:axe1_morphology} illustrate the microscopic origin of this broad boundary of outcomes. The three configurations were obtained at the same $L$, $J_2/J_1$, and cooling time, close to the kinetic boundary, yet they exhibit distinct final morphologies. The low-$M_f$ and intermediate-$M_f$ states are not simply random high-temperature states. They contain locally stripe-ordered regions separated by defects, domain walls, or region of competing stripe orientations, so the global stripe order remains weak or incomplete. The broad distribution of \(M_f\) near the boundary support a collective coarsening bottleneck, rather than a failure to form local stripe correlations.

The morphology is quantified using a local stripe-orientation field, whose algorithmic definition is given in Sec.~\ref{app:morphology_proxies}. At each site, a local orientation label is assigned from the strongest of the three local stripe-orientation scores, and connected components of equal orientation define local stripe domains. This construction provides practical real-space diagnostics rather than exact thermodynamic observables. The local scores, the connectivity rule, and the treatment of ambiguous sites are kept fixed for all parameters. The effective domain length $\ell_{\mathrm{eff}}$, the dominant-gage domain-wall proxy $\rho_{\mathrm{DW}}$, and the density of elementary triangular defects $\rho_{\triangle}^{\mathrm{defect}}$ are used. The quantity $\ell_{\mathrm{eff}}/L$ should be interpreted as a finite-size normalized morphology scale, not as an asymptotic correlation length.

For compact classification, a final configuration is called a single-domain stripe when it satisfies
\begin{equation}
M_{\max}\geq0.7,\qquad \rho_{\mathrm{DW}}\leq0.15.
\label{eq:single_domain_criterion}
\end{equation}
This criterion is deliberately stricter and more morphological than the global condition $M_f\geq0.5$ used to define the kinetic boundary in previous sections. It is used only as an operational label. A configuration that fails Eq.~\eqref{eq:single_domain_criterion} may still contain substantial local stripe correlations.

\begin{figure*}[t]
    \centering
    \includegraphics[width=\textwidth]{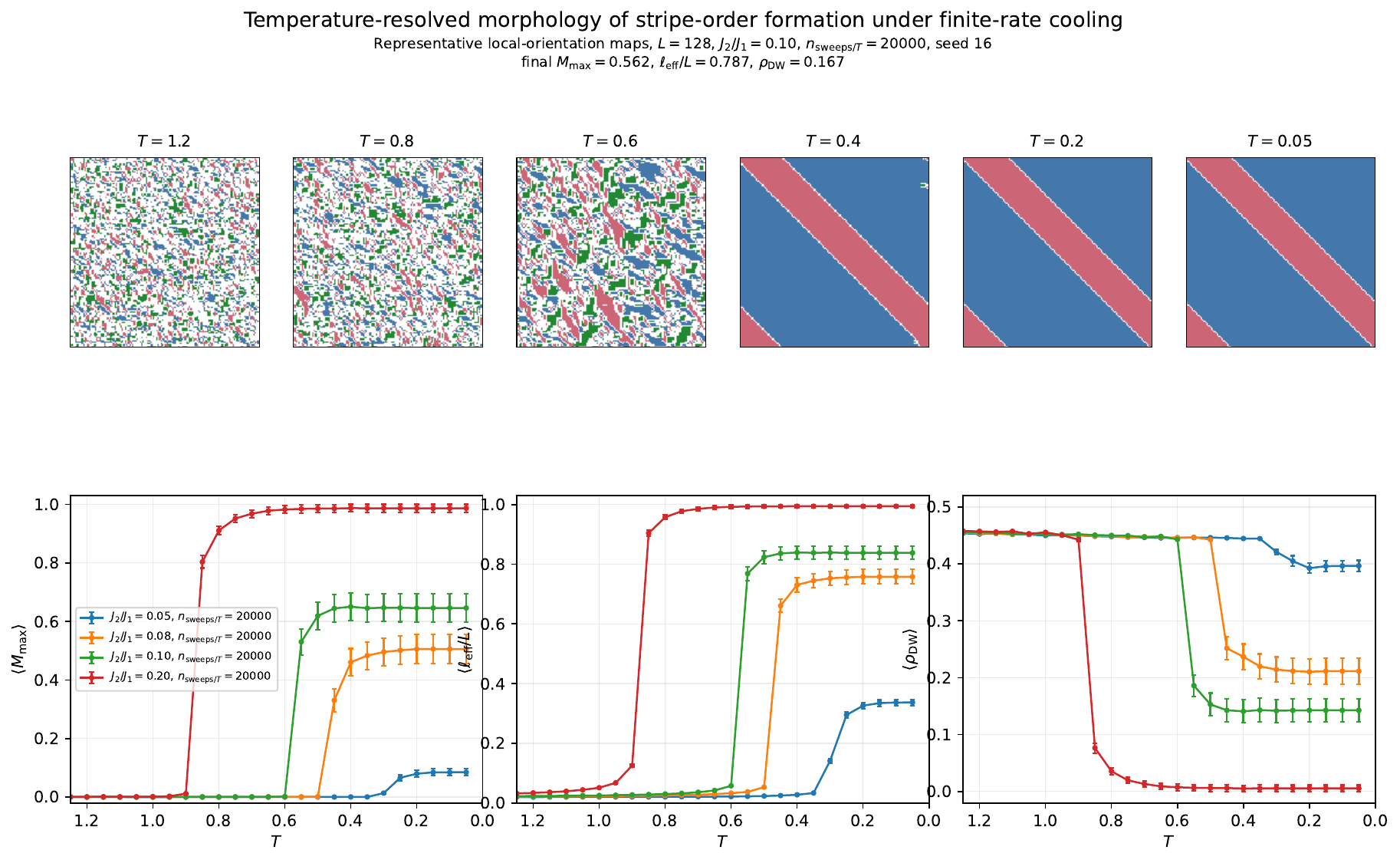}
    \caption{Temperature-resolved morphology of stripe-order formation at $L=128$. Top row: local stripe-orientation maps for $J_2/J_1=0.10$, $n_{\mathrm{sweeps}}/T=20000$, and seed 16; white pixels are locally ambiguous sites. This trajectory ends in a partially coarsened multidomain state with final $M_{\max}=0.562$, $\ell_{\mathrm{eff}}/L=0.787$, and $\rho_{\mathrm{DW}}=0.167$. Bottom row: ensemble-averaged evolution of $\langle M_{\max}\rangle$, $\langle\ell_{\mathrm{eff}}/L\rangle$, and $\langle\rho_{\mathrm{DW}}\rangle$ for $J_2/J_1=0.05,0.08,0.10,0.20$ at $n_{\mathrm{sweeps}}/T=20000$. Increasing $J_2/J_1$ sharpens stripe growth, increases the effective domain scale, and reduces the residual wall density.}
\label{fig:axis3_temperature_resolved_morphology}
\end{figure*}

Figure~\ref{fig:axis3_temperature_resolved_morphology} illustrates this morphology-based ordering pathway. It combines representative temperature-resolved local-orientation maps with ensemble-averaged curves for \(M_{\max}(T)\), \(\ell_{\mathrm{eff}}(T)/L\), and \(\rho_{\mathrm{DW}}(T)\), allowing the visual domain formation process to be connected with quantitative morphology observables.

The temperature-resolved maps show that local stripe orientations appear before the system has selected a unique global orientation. The ensemble averages show the same process quantitatively: increasing \(J_2/J_1\) makes the growth of \(M_{\max}\) and \(\ell_{\mathrm{eff}}/L\) sharper and reduces the residual wall density more efficiently. This supports the interpretation that the main bottleneck is not the initial appearance of local stripe correlations, but their subsequent coarsening into a single macroscopic orientation.

The morphology observables also provide a direct way to interpret the effective growth of the kinetic boundary. In a finite-rate coarsening picture, the boundary \(n^*(L,J_2/J_1)\) corresponds to the cooling time at which the characteristic stripe-domain scale becomes comparable to \(L\) for roughly half of the trajectories. This scale is represented operationally by $ell_{\mathrm{eff}}/L$ and \(\rho_{\mathrm{DW}}\). These quantities support the interpretation that failed trajectories are limited by incomplete domain coarsening rather than by the absence of local stripe correlations.

The clearest separation of regimes is obtained at $L=128$, where finite-size effects are strongest. Table~\ref{tab:axis3_L128_slow} summarizes the slowest available cooling results. For $J_2/J_1=0.05$, even increasing the cooling time to $40000$ sweeps per temperature value does not produce a single-domain stripe state. The global stripe order remains small, with $\langle M_{\max}\rangle=0.169$, and the largest local domain occupies only about $0.41$ of the locally valid sites. The effective domain length nevertheless increases under the slowest cooling conditions, reaching \(\ell_{\mathrm{eff}}/L\simeq0.48\), indicating the formation of sizeable local stripe regions without global orientation selection.

At \(J_2/J_1=0.08\) and \(0.10\), slower cooling produces progressively larger stripe domains and lower residual wall density, but single-domain selection remains probabilistic at \(L=128\). By contrast, \(J_2/J_1=0.20\) reaches an almost system-spanning stripe domain within the simulated window.

\begin{table*}[t]
\centering
\small
\begin{tabular}{c c c c c c}
\toprule
$J_2/J_1$ & $n_{\mathrm{sweeps}}/T$ & $\langle M_{\max}\rangle$ & $\langle \ell_{\mathrm{eff}}/L\rangle$ & $\langle \rho_{\mathrm{DW}}\rangle$ & $P_{\mathrm{single}}$ \\
\midrule
0.05 & 40000 & 0.169 & 0.480 & 0.351 & 0.00 \\
0.08 & 40000 & 0.608 & 0.819 & 0.165 & 0.40 \\
0.10 & 40000 & 0.768 & 0.895 & 0.096 & 0.62 \\
0.20 & 20000 & 0.987 & 0.994 & 0.006 & 0.98 \\
\bottomrule
\end{tabular}
\caption{Representative final morphology at $L=128$ in the slowest available cooling runs. The single-domain probability is based on Eq.~\eqref{eq:single_domain_criterion}. The table reports seed averages; standard errors are shown in the corresponding morphology figures when plotted.}
\label{tab:axis3_L128_slow}
\end{table*}

These data separate three finite-time morphological regimes. At weak next-nearest-neighbor coupling, represented here by $J_2/J_1=0.05$, the system develops local stripe correlations but remains a robust mosaic on the accessible cooling times. At intermediate coupling, $J_2/J_1=0.08$ and $0.10$, the system forms large domains and can sometimes reach a single-domain stripe state, but global orientation selection remains probabilistic and strongly rate-dependent at $L=128$. At stronger coupling, $J_2/J_1=0.20$, the same local dynamics reaches a nearly system-spanning stripe domain for sufficiently slow cooling.

A useful distinction emerges by comparing the triangular-defect density with the residual domain-wall density. In the slow $L=128$ runs, $\rho_{\triangle}^{\mathrm{defect}}$ is essentially zero for the weak and intermediate couplings once the system has reached low temperatures. For example, at $J_2/J_1=0.05$ and $40000$ sweeps per temperature value, the triangular-defect density is zero within the measured resolution, while $\rho_{\mathrm{DW}}\simeq0.351$. Similarly, at $J_2/J_1=0.08$ and $40000$ sweeps per temperature value, $\rho_{\triangle}^{\mathrm{defect}}\simeq0$ but $\rho_{\mathrm{DW}}\simeq0.165$. Therefore, the residual disorder in the final low-temperature states is not dominated by elementary triangles with three equal spins. Instead, the system has largely eliminated local nearest-neighbor constraint violations but remains morphologically heterogeneous because different locally stripe-ordered domains have not merged into a unique global orientation.

This distinction is important for the physical interpretation. If the failure to order were mainly due to a finite density of local thermal defects, one would expect the low-temperature state to substantially improve as $\rho_{\triangle}^{\mathrm{defect}}$ vanishes. The data show the opposite: local defects can disappear while domain walls and orientation mismatches persist. The bottleneck is therefore a late-stage coarsening and orientation-selection process rather than the initial formation of local stripe correlations. This is consistent with theoretical descriptions in which the stripe phase and its disordering are governed by domain walls and by the topological structure of the nearby nearest-neighbor manifold \cite{Korshunov2005,Smerald2016}.

Thus, finite-rate cooling proceeds through local stripe formation followed by domain coarsening, while global orientation selection remains incomplete when the cooling is too fast, the system is too large, or \(J_2/J_1\) is too weak.

\subsection{Local constraint does not imply global stripe selection}

In the weak-\(J_2/J_1\) crossover, the energy criterion can be satisfied before the order-parameter criterion. The finite-time bottleneck cannot be reduced to energetic relaxation alone. The system may approach the stripe energy while retaining residual domain walls or competing stripe orientations.

This separation shows that the crossover is not purely energetic: $J_2/J_1$ first biases the system toward low-energy configurations, while global orientation selection remains limited by slow domain-wall and sector dynamics. The broad distribution of $M_f$ reflects the coexistence of nearly ordered and multidomain outcomes across cooling trajectories.”

The triangular-defect density remains negligible ($\leq10^\mathrm{−5}$ -- $10^\mathrm{−6}$) in the refined weak-$J_2/J_1$ data, including at large $L$, showing that the blocked states remain close to the nearest-neighbor constrained manifold.

This result gives the weak-$J_2/J_1$ regime its specific physical meaning. For $J_2/J_1\leq0.04$, and still largely for $J_2/J_1=0.050$--$0.055$ at large sizes, the system cools into states that satisfy the local nearest-neighbor constraint almost perfectly but do not globally select the stripe ground state. The system has therefore restored the local nearest-neighbor constraint to numerical precision before the weak perturbation produced complete global stripe selection. This behavior is consistent with the known dimer, string, and height descriptions of the TLIAF, in which the constrained manifold is partitioned into sectors that are not freely connected by local moves~\cite{BloteHilhorst1982,BloteNightingale1993,Kim2007,Smerald2016}.

\subsection{A mobility window for late-stage coarsening}

The fixed-temperature holds identify a mobility window in which the stripe phase remains stable while the defects remain mobile enough for coarsening. Below this window, the stripe state is stable, but random initial conditions remain trapped in multidomain states. Above this window, stripe order is progressively destabilized. The finite-rate cooling failures can therefore be interpreted as trajectories that cross the mobility window too quickly and become kinetically trapped before global orientation cleanup is complete.

The most detailed cooling-rate scan was performed for $L=64$ and $J_2/J_1=0.08$, using 100 independent seeds. The values reported in Table~\ref{tab:cooling_rate_J208} show that the different observables respond differently to the cooling rate. Already at $n_{\mathrm{sweeps}}/T=1000$, the average dominant stripe order parameter is relatively high, $\langle M_{\max}\rangle=0.705$, indicating that substantial stripe order has already developed. However, the strict near-ground-state success probability is only $P_{\mathrm{stripe}}^{\mathrm{strict}}=0.50$, while the residual domain-wall proxy and the orientation-competition parameter remain substantial, with $\langle\rho_{\rm DW}\rangle=0.195$ and $\langle C_{\rm orient}\rangle=0.250$. This shows that fast cooling does not prevent the formation of local stripe order; rather, it limits the late-stage removal of domain walls and the global selection of a single stripe orientation.

When the cooling is slowed down, $P_{\mathrm{stripe}}^{\mathrm{strict}}$ increases from $0.50$ at $n_{\mathrm{sweeps}}/T=1000$ to $0.96$ at $n_{\mathrm{sweeps}}/T=20000$. Over the same range, $\langle M_{\max}\rangle$ increases from $0.705$ to $0.975$, but the most pronounced evolution occurs in the defect-related observables: $\langle \rho_{\rm DW}\rangle$ decreases from $0.195$ to $0.016$, and $\langle C_{\rm orient}\rangle$ decreases from $0.250$ to $0.025$. Thus, slowing the cooling does not merely increase the amount of stripe order; it also promotes the transformation of a multidomain stripe-like configuration into a nearly single-domain stripe state. The increase of $\langle W_{\max}^{\rm proxy}\rangle$ from $45.13$ to $62.42$, close to the ideal value $L=64$, supports the same interpretation.

These trends suggest a two-stage ordering process. A substantial dominant stripe order can develop while multiple spatial domains with competing orientations are still present. The slower process is the global coarsening stage, during which competing stripe orientations are progressively eliminated and domain walls are removed. The failures observed at faster cooling rates are therefore primarily kinetic: the system can still form stripe order locally, but it does not have enough time to reach a clean global stripe orientation before the dynamics become too slow.

The strength of $J_2/J_1$ controls the energetic bias of the stripe phase. Comparing different values of $J_2/J_1$ at fixed system size shows that global stripe formation becomes easier as $J_2/J_1$ increases. For $L=64$, the simulations indicate that $J_2/J_1=0.15$ reaches the stripe state readily, $J_2/J_1=0.10$ raches it for intermediate cooling rates, while $J_2/J_1=0.08$ is the most difficult case among the tested values.

The difficult regime is therefore associated with the weak stripe selection rather than with large $J_2/J_1$. When $J_2/J_1$ is small but nonzero, the stripe phase is energetically selected, yet the driving force toward stripe order is weak. The system can then enter a low-temperature regime before domain coarsening is complete, leading to kinetically trapped multidomain configurations.

To distinguish kinetic trapping from simple thermal destabilization of the stripe state, fixed-temperature hold simulations were performed from two different initial conditions. The first was a random configuration of the high-temperature disordered state. The second is a perfect stripe initial configuration. If both initial conditions converge to the same state at a given temperature, this indicates efficient relaxation on the simulated timescale. If the perfect stripe remains stable while the random initial condition remains only partially ordered, this initial-condition dependence is consistent with slow coarsening and kinetic trapping. The \(J_2/J_1=0.08\) case is summarized in Table~\ref{tab:metastability_J208}. These data show that the stripe state remains stable at these temperatures, whereas random initial conditions can remain trapped in multidomain configurations.

At higher temperatures the random branch becomes much more efficient at reaching the stripe state. Around $T=0.40$, the random initial condition reaches
\begin{equation}
P_{\mathrm{stripe}}^{\mathrm{strict}}\simeq 0.90,
\end{equation}
and around $T=0.45$ it reaches
\begin{equation}
P_{\mathrm{stripe}}^{\mathrm{strict}}=1.
\end{equation}
This identifies an mobility window around $T\simeq 0.40$ -- $0.45$ for $J_2/J_1=0.08$. In this range, the stripe phase is still sufficiently stable, but domain walls and defects are mobile enough to allow coarsening toward a single-orientation stripe state.

For $J_2/J_1=0.10$, the same mechanism is observed, but the kinetic trapping is weaker. The perfect stripe state remains stable from approximately $T=0.40$ to $T=0.55$. Random initial conditions remain partially blocked at $T=0.40$ and $T=0.45$, but reach the stripe state for $T=0.50$ -- $0.55$. At $T=0.60$, both branches become disordered. Thus, the mobility window is shifted to higher temperatures and the accessibility of the stripe state is improved compared with $J_2/J_1=0.08$.

For $J_2/J_1=0.15$, the random and perfect-stripe branches become almost indistinguishable in the ordered region. The stripe phase is much more easily accessible from random initial conditions. At $T=0.70$, $M_{\max}$ remains high, around $0.96$, although the strict stripe-success probability can decrease because the energy criterion is increasingly difficult to satisfy at finite temperature. At $T=0.80$, both branches become disordered. This supports the interpretation that a stronger $J_2/J_1$ reduces the severity of kinetic trapping by stabilizing stripe order at temperatures where relaxation remains efficient.

The fixed-temperature holds do not primarily reveal a wide thermodynamic coexistence window between two stable equilibrium phases. Instead, they show a strong dependence on the initial basin of attraction and the mobility of defects. A perfect stripe state remains stable deep in the ordered region, while a random state can remain trapped in a multidomain configuration if the temperature is too low for domain walls to move efficiently.

For $J_2/J_1=0.08$, this window is centered around $T\simeq 0.40$ -- $0.45$, while for $J_2/J_1=0.10$ it shifts to higher temperatures and trapping is weaker. This provides a dynamical interpretation of the finite-rate boundary: a cooling trajectory can pass through the regime in which domain-wall motion is efficient without spending enough time there to eliminate competing stripe domains.

\begin{figure}[t]
\centering
\includegraphics[width=0.5\textwidth]{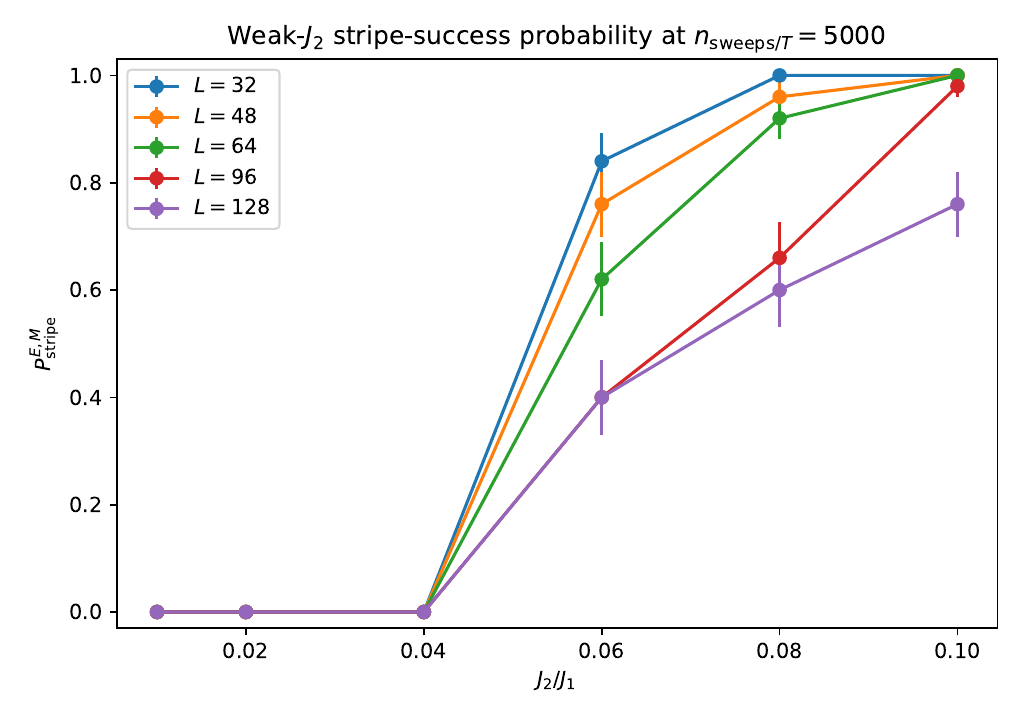}
\caption{Weak-coupling stripe-success probability $P_{\mathrm{stripe}}^{E,M}$ at $n_{\mathrm{sweeps}}/T=5000$ as a function of $J_2/J_1$ for $L=32,48,64,96,128$. The success criterion combines the energy condition $e_f-e_{\mathrm{stripe}}<0.1\,J_2/J_1$ with the order-parameter condition $M_f>0.75$. The probability remains zero up to $J_2/J_1=0.04$ for all sizes, rises around $J_2/J_1\simeq0.06$, and the crossover shifts toward larger $J_2/J_1$ as $L$ increases.}
\label{fig:axis4_Pstripe_s5000}
\end{figure}

Figure~\ref{fig:axis4_Pstripe_s5000} and Table~\ref{tab:axis4_Pstripe_s5000} summarize the slowest coarse cooling rate in the coarse scan, $n_{\mathrm{sweeps}}/T=5000$. The suppression of strict stripe success at the weakest couplings is pronounced. For $J_2/J_1\leq0.04$, the strict stripe-success probability remains zero for every simulated size. Thus, within the present local-dynamics protocol and cooling window, weak antiferromagnetic next-nearest-neighbor couplings in this range do not lead to global stripe selection, even though the stripe state is the energetically selected ground state for any $J_2/J_1>0$.

The size dependence is also clear. At fixed $J_2/J_1=0.06$, the stripe-success probability decreases with increasing $L$, showing that the weak-$J_2/J_1$ crossover is also strongly size-dependent. Hence, the weak-$J_2/J_1$ crossover is controlled not only by the perturbation strength but also by the system size. Larger systems, therefore, require more time to achieve global stripe selection within the same cooling protocol.

The mean stripe order parameter provides a second view of the same process. The mean stripe-order values remain relatively small and decrease with increasing system size, consistent with the absence of global stripe selection in the weak coupling regime. By contrast, $\langle M_f\rangle$ rises for higher coupling, but the strict success probability remains below one at $L=128$. This difference between mean order and strict success already indicates that a substantial fraction of trajectories are partially stripe-ordered but not clean single-orientation stripe states.

Together, these results support a two-stage ordering pathway. Local stripe correlations and the nearest-neighbor constraint are restored relatively rapidly, whereas the subsequent elimination of domain walls and competing stripe orientations is substantially slower. Fast cooling, therefore, leaves the system kinetically trapped after local stripe formation but before global stripe selection is complete.

\section{Weak-$J_2$ regime: energetic selection without global stripe order}
We now examine how this kinetic bottleneck changes as the energetic bias $J_2/J_1$ is reduced.

The combined coarse and refined data support a three-regime picture. For $J_2/J_1\leq0.04$, the perturbation too weak to produce global stripe order within the simulated time scales. The final states contain very few local constraint violations and remain close to the nearest-neighbor constrained manifold, while $P_{\mathrm{stripe}}^{E,M}=0$ throughout the coarse grid. For $J_2/J_1\simeq0.05$--$0.07$, the system enters a crossover regime. The mean stripe order and the probability of satisfying the energy criterion increase, but the outcome remains strongly seed-dependent, and the strict stripe-success probability remains well below unity at large $L$. For $J_2/J_1\gtrsim0.08$ on the coarse grid, global stripe states become dynamically accessible for small and intermediate sizes, although substantial finite-time failures remain at the largest sizes.

For $L=96$ -- $128$ and $n_{\mathrm{sweeps}}/T=5000$, the operational crossover for the chosen finite-time criterion occurs around $J_2/J_1\simeq0.065$ -- $0.073$. In other words, $J_2/J_1\simeq0.05$ marks the onset of an energetic and partial orientational bias, whereas robust single-orientation stripe selection requires a somewhat larger perturbation or a slower cooling schedule.

The energy criterion is satisfied more readily than the global stripe criterion in the weak-\(J_2/J_1\) regime, showing that the system can approach the low-energy manifold without selecting a unique stripe orientation. The fitted \(J_c^{\mathrm{kin}}\simeq0.058\) should therefore be regarded as an effective finite-window onset scale, not as a sharp
threshold for global stripe accessibility.

\section{Conclusions}\label{sec:conclusion}

We have studied the finite-rate cooling dynamics of the triangular-lattice $J_1$ -- $J_2$ Ising antiferromagnet under local single-spin Metropolis dynamics. For $J_2/J_1>0$, the antiferromagnetic next-nearest neighbor coupling selects a collinear stripe ground state, but the simulations show that this thermodynamic selection does not automatically imply finite-time dynamical accessibility. The main quantitative diagnostic is an operational stripe-accessibility scale $n^*(L,J_2/J_1)$, defined by $P(M_f\geq0.5)=0.5$. This accessibility scale increases strongly with system size and as $J_2/J_1$ is reduced. The empirical form $A L^\mathrm{z_eff}[(J_2/J_1)-J_c^{\mathrm{kin}}]^{-\alpha}$ is used only as a finite-window parameterization of this operational boundary, rather than as an asymptotic scaling law.

The main physical result is that failed cooling trajectories are not generally high-energy disordered states. They often contain locally stripe-ordered regions and have largely restored the nearest-neighbor triangular antiferromagnetic constraint, but retain residual domain walls and competing stripe orientations. Fixed-temperature holds show that prepared stripe states can remain stable while random initial conditions remain trapped in multidomain configurations. The limitation is therefore mainly kinetic: local stripe correlations form before the system has completed the slower removal of domain walls and the global selection of a single stripe orientation.

The real-space morphology and weak-$J_2/J_1$ analyses support the same interpretation. Morphological diagnostics show that stripe order develops through domain formation and coarsening rather than through homogeneous ordering. In the weak-coupling regime, the final energy can approach the stripe ground-state energy before the global stripe order criterion is satisfied. The near-zero triangular-defect density further indicates that the system can return close to the nearest-neighbor constrained manifold without necessarily selecting a single stripe sector. Thus, energetic degeneracy lifting, local constraint restoration, and global stripe-sector selection occur on distinct finite-time scales.

The present conclusions are restricted to local Metropolis dynamics, finite system sizes, and the cooling schedules studied here. The quantities $n^*(L,J_2/J_1)$, $z$, $J_c^{\mathrm{kin}}$ and $\alpha$ should therefore be interpreted as protocol-dependent accessibility measures, not as equilibrium phase-boundary parameters. The present data allow a comparison between finite-rate energy relaxation, stripe-order growth, and domain-wall removal, but they do not constitute equilibrium thermodynamic sampling.

Thus, under finite-rate local cooling, energetic relaxation and restoration of the local frustrated constraint accompany the formation of stripe domains, while the subsequent coarsening and global stripe-orientation selection remain substantially slower. The ordering process becomes dynamically inaccessible primarily at this late stage when the cooling is too fast or $J_2/J_1$ is too weak.

{\bf Acknowledgements:} The author thanks Yves Lansac for his supervision and guidance, Stam Nicolis of the Institut Denis Poisson and Pascal Thibaudeau of the CEA/Le Ripault for discussions, suggestions, and comments on the manuscript. The authors also thanks Vinh Ta-phuoc for his supervision and the GREMAN for autorising the use of their calculation machines.

\clearpage
\appendix
\makeatletter
\renewcommand{\thesection}{\Alph{section}}
\renewcommand{\thesubsection}{\thesection.\arabic{subsection}}
\renewcommand{\p@subsection}{}
\renewcommand{\thetable}{\thesection\arabic{table}}
\renewcommand{\theHtable}{\thesection.\arabic{table}}
\@addtoreset{table}{section}
\makeatother

\section{Numerical validation, morphology definitions, and statistical conventions}
\label{app:numerical_details}

\subsection{Numerical validation checks}

The triangular lattice is represented by integer coordinates \((x,y)\), with \(0\leq x,y<L\) and periodic boundary conditions. The nearest-neighbor offsets are
\begin{equation}
(\pm1,0),\quad (0,\pm1),\quad (1,-1),\quad (-1,1),
\end{equation}
and the next-nearest-neighbor offsets are
\begin{equation}
(1,1),\quad (-1,-1),\quad (2,-1),\quad (-2,1),\quad (1,-2),\quad (-1,2).
\end{equation}
Production runs use precomputed neighbor tables. The colored update schedule partitions the lattice into \(4\times4=16\) color classes according to \((x\bmod 4,y\bmod 4)\). Sites in the same color class do not interact through the nearest-neighbor or next-nearest-neighbor bonds of Eq.~\eqref{eq:hamiltonian}, so one color class can be updated without internal conflicts.

Several deterministic configurations were used to validate the energy and order-parameter conventions. For a ferromagnetic configuration, the expected energy per spin is
\begin{equation}
e_{\mathrm{FM}}=3J_1+3J_2,
\end{equation}
because there are three nearest-neighbor and three next-nearest-neighbor bonds per site. The validation also checks that \(M_{\mathrm{stripe}}=1\) for each of the three stripe orientations, that the triangular-defect density vanishes in perfect stripe states, and that the fraction of frustrated nearest-neighbor bonds is \(1/3\). Random-state checks give \(\langle \rho_{\triangle}^{\mathrm{defect}}\rangle\simeq1/4\), \(\langle f_{\mathrm{NN}}^{\mathrm{frustrated}}\rangle\simeq1/2\), and small \(M_{\mathrm{stripe}}\), as expected.

\subsection{effective fit of $n^*$}\label{sub;effective_fit}

The measured $n^*$ increases strongly with system size. Over the simulated range, the data at $J_2/J_1=0.10$ are compatible with an approximately $L^2$ like growth, while the weaker coupling data show stronger deviations and broader crossover behavior. We therefore use the empirical form
\begin{equation}
n^*(L,J_2/J_1) = A L^\mathrm{z_eff} \left(\frac{J_2}{J_1}-J_c^{\mathrm{kin}}\right)^{-\alpha}.
\label{eq:kinetic_fit_form}
\end{equation}
only as a finite-window parameterization rather than as an asymptotic scaling law. The corresponding fits and uncertainties are given in Table~\ref{tab:axe1_fit_extended}, and Table~\ref{tab:axe1_mechanism}.

This interpretation is consistent with the general theory of phase-ordering kinetics, in which ordering after a quench proceeds through the growth of a characteristic domain scale controlled by the motion and annihilation of defects or domain walls~\cite{Bray1994}.

\subsection{Definition of the morphology proxies}\label{app:morphology_proxies}

This subsection gives the explicit algorithm used to extract the real-space morphology observables from a spin configuration. Sites are labeled by integer coordinates \(\mathbf r=(x,y)\), with periodic boundary conditions. The three nearest-neighbor axes are:
\begin{equation}
\mathbf e_0=(1,0),\qquad \mathbf e_1=(0,1),\qquad \mathbf e_2=(1,-1),
\end{equation}
so that the six nearest neighbors of a site are \(\mathbf r\pm\mathbf e_a\), with \(a=0,1,2\).

For each site \(\mathbf r\), the number of equal-spin nearest neighbors along each axis is first counted,
\begin{equation}
c_a(\mathbf r)=\left[s_{\mathbf r}=s_{\mathbf r+\mathbf e_a}\right]+\left[s_{\mathbf r}=s_{\mathbf r-\mathbf e_a}\right],
\qquad a=0,1,2.
\end{equation}
Here \([\cdots]\) is the Iverson Bracket~\cite{Graham1994}, if the condition inside is true, it is equal to one, else it is null. In a perfect stripe state, one of the three axes has \(c_a=2\), while the other two axes have \(c_a=0\). The local orientation label \(o(\mathbf r)\) is then assigned by
\begin{equation}
o(\mathbf r)=
\begin{cases}
\arg\max_a c_a(\mathbf r), & \begin{gathered} c_{\max}(\mathbf r)\geq2,\\ c_{\max}(\mathbf r)-c_{\mathrm{2nd}}(\mathbf r)\geq1, \end{gathered}\\
-1, & \mathrm{otherwise}.
\end{cases}
\end{equation}
Here \(c_{\max}\) is the largest of the three values \(c_0,c_1,c_2\), and \(c_{\mathrm{2nd}}\) is the second largest. The value \(o=-1\) denotes a locally ambiguous or non-stripe-like site. In the simulations reported in the main text, the thresholds are therefore \(c_{\max}\geq2\) and \(c_{\max}-c_{\mathrm{2nd}}\geq1\).

The set of locally valid sites is
\begin{equation}
\mathcal V=\{\mathbf r\mid o(\mathbf r)\geq0\},
\end{equation}
with \(N_{\mathcal V}=|\mathcal V|\). Connected components are then computed on the graph of valid sites, using nearest-neighbor connectivity on the triangular lattice. Two valid sites \(\mathbf r\) and \(\mathbf r'\) are in the same component if they are nearest neighbors and have the same local orientation label,
\begin{equation}
o(\mathbf r)=o(\mathbf r')\geq0.
\end{equation}
Let \(A_k\) be the area, in number of sites, of component \(k\). The largest-domain fraction used in the main text is defined as
\begin{equation}
f_{\mathrm{largest}}=\frac{\max_k A_k}{N_{\mathcal V}},
\end{equation}
with \(f_{\mathrm{largest}}=0\) if \(N_{\mathcal V}=0\). This definition normalizes the largest domain by the number of locally valid sites, not by the total number of lattice sites.

The effective domain area is defined by the second moment of the component-size distribution,
\begin{equation}
A_{\mathrm{eff}}=\frac{\sum_k A_k^2}{\sum_k A_k}=\frac{\sum_k A_k^2}{N_{\mathcal V}},
\end{equation}
and the effective domain length is
\begin{equation}
\ell_{\mathrm{eff}}=\sqrt{A_{\mathrm{eff}}}.
\end{equation}
This quantity is biased toward large domains. It is therefore a practical finite-size morphology scale rather than an asymptotic correlation length. In the main text the normalized quantity is \(\ell_{\mathrm{eff}}/L\).

The domain-wall proxy \(\rho_{\mathrm{DW}}\) used in the morphology figures and tables is computed independently of the local connected-component map. It is first determined by the dominant global stripe orientation
\begin{equation}
\hat a=\arg\max_{a=0,1,2} M_a,
\end{equation}
where \(M_a\) are the three global stripe structure-factor components. For each orientation, a reference stripe parity field is defined:
\begin{equation}
\begin{aligned}
p_0(x,y)&=(-1)^x,\\
p_1(x,y)&=(-1)^y,\\
p_2(x,y)&=(-1)^{x+y}.
\end{aligned}
\end{equation}
The spin configuration is then transformed into the gage of the dominant stripe orientation by
\begin{equation}
\tau_{\mathbf r}=s_{\mathbf r}p_{\hat a}(\mathbf r).
\end{equation}
A perfect stripe state of orientation \(\hat a\), including its globally spin-flipped partner, has uniform \(\tau_{\mathbf r}\). The domain-wall proxy is the fraction of nearest-neighbor bonds on which this gaged variable changes sign,
\begin{equation}
\rho_{\mathrm{DW}}=\frac{1}{3L^2}\sum_{\mathbf r}\sum_{a=0}^{2}\mathbb{1}\!\left[\tau_{\mathbf r}\neq\tau_{\mathbf r+\mathbf e_a}\right].
\end{equation}
Only the three positive bond directions are counted, so each nearest-neighbor bond is counted once. With this normalization, \(\rho_{\mathrm{DW}}=0\) indicates a perfect single-orientation stripe state, while larger values indicate residual walls, phase slips, or regions incompatible with the dominant global stripe gage.

The single-domain classification used in the morphology analysis is therefore
\begin{equation}
M_{\max}\geq0.7,\qquad f_{\mathrm{largest}}\geq0.7,\qquad \rho_{\mathrm{DW}}\leq0.15.
\end{equation}
This criterion is operational and protocol-dependent. It is not an equilibrium phase criterion; it is used only to distinguish nearly system-spanning stripe configurations from multidomain or mosaic-like final states.

\subsection{Statistical uncertainty estimates}\label{app:uncertainty_estimates}

All statistical uncertainties are estimated from the ensemble of independent cooling trajectories. For an observable \(X\) measured at fixed \(L\), \(J_2/J_1\), and \(n_{\mathrm{sweeps}}/T\), the reported mean is
\begin{equation}
\overline X=\frac{1}{N_{\mathrm{seeds}}}\sum_{s=1}^{N_{\mathrm{seeds}}}X_s,
\end{equation}
and error bars for continuous observables in the morphology plots are standard errors over seeds,
\begin{equation}
\mathrm{SEM}(X)=\frac{\sigma_X}{\sqrt{N_{\mathrm{seeds}}}}.
\end{equation}
These error bars quantify trajectory-to-trajectory fluctuations for the specified protocol. They should not be interpreted as equilibrium statistical errors from a single equilibrated Markov chain.

For binary success probabilities, such as \(P_{\mathrm{stripe}}\), \(P_{\mathrm{stripe}}^{\mathrm{strict}}\), \(P_{\mathrm{stripe}}^{E,M}\), and \(P_{\mathrm{single}}\), the central estimate is \(k/n\), where \(k\) is the number of successful trajectories among \(n=N_{\mathrm{seeds}}\) independent runs. Confidence intervals are estimated with Wilson intervals rather than with the naive binomial standard error, because the latter vanishes artificially when \(k=0\) or \(k=n\). For a confidence level corresponding to the normal quantile \(z\), the Wilson center and half-width are
\begin{equation}
p_{\mathrm W}=\frac{\hat p+z^2/(2n)}{1+z^2/n},
\end{equation}
\begin{equation}
\Delta p_{\mathrm W}=\frac{z}{1+z^2/n}\sqrt{\frac{\hat p(1-\hat p)}{n}+\frac{z^2}{4n^2}},
\end{equation}
where \(\hat p=k/n\). Unless otherwise stated, \(z=1.96\) is used for \(95\%\) Wilson intervals. For reference, near \(\hat p=0.5\), the \(95\%\) Wilson half-width is approximately \(0.14\) for \(n=50\), \(0.10\) for \(n=100\), and \(0.09\) for \(n=125\).

The kinetic boundary \(n^*(L,J_2/J_1)\) is a derived quantity and is therefore assigned an uncertainty by bootstrap resampling over seeds. For each bootstrap sample, the run-level data are resampled with replacement at fixed \(L\), \(J_2/J_1\), and \(n_{\mathrm{sweeps}}/T\). The success probability \(P_{\mathrm{stripe}}(n_{\mathrm{sweeps}}/T)\) is recomputed, isotonic regression is applied to enforce monotonicity in \(n_{\mathrm{sweeps}}/T\), and the crossing \(P_{\mathrm{stripe}}=0.5\) is re-extracted. The quoted central value is the value obtained from the original ensemble, while the uncertainty shown in the boundary figures is obtained from the bootstrap distribution of \(n^*\). Unless an explicit command-line value was used, \(N_{\mathrm{boot}}=100\) bootstrap resamples were used; command-line production runs used \(N_{\mathrm{boot}}=200\) when specified.

The empirical fit parameters in Eq.~\eqref{eq:kinetic_fit_form} are obtained from the central boundary estimates. Their uncertainty is assessed by repeating the fit on bootstrap-resampled boundary datasets. Because the available size and coupling windows are limited, these fit uncertainties quantify the stability of the finite-rate parameterization within the simulated dataset; they do not represent uncertainty on an asymptotic universal exponent.

For compactness, the main tables report central estimates only, except when a value is censored by the simulated sweep range. In such cases, inequalities indicate that the crossing lies outside the sampled \(n_{\mathrm{sweeps}}/T\) window. The corresponding error bars or bootstrap ranges are shown in the figures when they are resolved on the plotting scale.

\subsection{Finite-rate energy-derivative diagnostic}

As an additional finite-rate diagnostic, the numerical derivative \(d\langle e\rangle/dT\) was also computed along the cooling trajectories. This quantity is not an equilibrium specific heat, because the configurations are generated during a finite-rate protocol rather than by independent equilibrium sampling at fixed temperature. It is used only as a heat-release proxy indicating the temperature window over which the energy changes most rapidly.

\begin{figure}[t]
\centering
\includegraphics[width=0.48\textwidth]{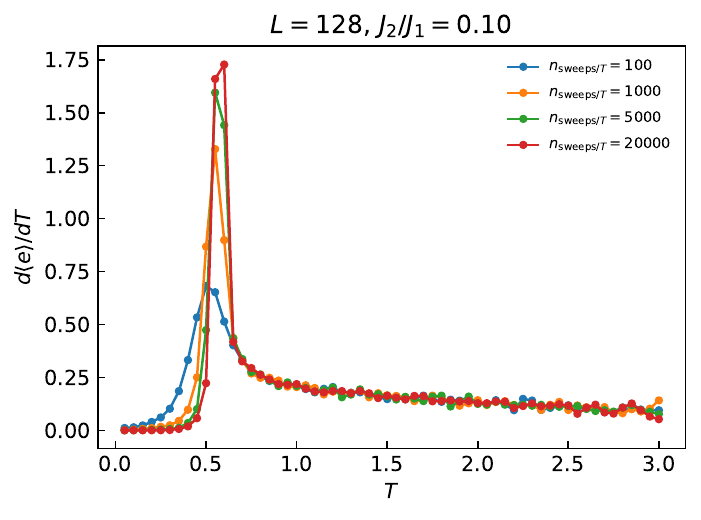}
\caption{Finite-rate energy-derivative signal \(d\langle e\rangle/dT\) for \(L=128\), \(J_2/J_1=0.10\), and several cooling rates. This quantity is a nonequilibrium heat-release proxy along the cooling trajectories, not an equilibrium specific heat.}
\label{fig:dynamic_heat_proxy}
\end{figure}

\subsection{Hold temperatures}

\begin{table}[!htbp]
\centering
\small
\begin{tabular}{c c c c c}
\toprule
Sweeps & \(P_{\mathrm{stripe}}^{\mathrm{strict}}\) & \(\langle M_{\max}\rangle\) & \(\langle\rho_{\rm DW}\rangle\) & \(\langle C_{\rm orient}\rangle\) \\
\midrule
\(10^5\) & 0.367 & 0.673 & 0.210 & 0.311 \\
\(3\times10^5\) & 0.760 & 0.890 & 0.0736 & 0.110 \\
\bottomrule
\end{tabular}
\caption{Effect of hold time at \texorpdfstring{$L=96$}{L=96}, \texorpdfstring{$J_2/J_1=0.08$}{J2/J1=0.08}, and \texorpdfstring{$T=0.40$}{T=0.40}.}
\label{tab:hold_time_L96}
\end{table}
The strong improvement between \(10^5\) and \(3\times10^5\) sweeps provides direct evidence that the trapping is largely kinetic. Many trajectories that remain blocked after \(10^5\) sweeps eventually reach a nearly ordered stripe state when the hold time is increased. However, the persistence of failed runs after \(3\times10^5\) sweeps shows that the relaxation time at \(L=96\) remains large.

\subsection{Threshold robustness}

To verify that the weak-\(J_2/J_1\) conclusion does not depend on a single arbitrary success threshold, the run-level data were reanalyzed using several order-parameter and energy tolerances. A trajectory was classified as successful when
\begin{equation}
e_f-e_{\mathrm{stripe}} < \epsilon_E,
\end{equation}
\begin{equation}
M_f > M_{\mathrm{th}},
\end{equation}
with
\begin{equation}
M_{\mathrm{th}}\in\{0.60,0.70,0.75,0.80\},
\end{equation}
\begin{equation}
\epsilon_E/(J_2/J_1)\in\{0.05,0.10,0.20\}.
\end{equation}
For each pair of thresholds, the probability \(P_{\mathrm{stripe}}^{E,M}\) and the interpolated value of \((J_2/J_1)^\star\) at which \(P_{\mathrm{stripe}}^{E,M}=0.5\) were recomputed. The qualitative conclusion is unchanged: the regime \(J_2/J_1\leq0.04\) does not dynamically reach a global stripe state in the simulated time window, while the large-size crossover remains in the range \(J_2/J_1\simeq0.06\text{--}0.07\), with the precise value depending on the strictness of the criterion.

\begin{table}[!htbp]
\centering
\small
\begin{tabular}{c c c c}
\toprule
$M_{\mathrm{th}}$ & $\epsilon_E/(J_2/J_1)$ & $L=96$ & $L=128$ \\
\midrule
0.60 & 0.05 & 0.059 & 0.060 \\
0.60 & 0.10 & 0.057 & 0.059 \\
0.60 & 0.20 & 0.057 & 0.059 \\
0.70 & 0.05 & 0.063 & 0.063 \\
0.70 & 0.10 & 0.062 & 0.060 \\
0.70 & 0.20 & 0.062 & 0.060 \\
0.75 & 0.05 & 0.071 & 0.073 \\
0.75 & 0.10 & 0.071 & 0.073 \\
0.75 & 0.20 & 0.071 & 0.073 \\
0.80 & 0.05 & 0.073 & 0.076 \\
0.80 & 0.10 & 0.073 & 0.076 \\
0.80 & 0.20 & 0.073 & 0.076 \\
\bottomrule
\end{tabular}
\caption{Robustness of the weak-$J_2/J_1$ crossover estimate at $n_{\mathrm{sweeps}}/T=5000$. The entries give the isotonic interpolation of $\left(J_2/J_1\right)^\star$ defined by $P_{\mathrm{stripe}}^{E,M}=0.5$ for different order-parameter thresholds $M_{\mathrm{th}}$ and energy tolerances $\epsilon_E/(J_2/J_1)$.}
\label{tab:axis4_threshold_robustness}
\end{table}

The crossover estimate is most sensitive to the order threshold $M_{\mathrm{th}}$ and only weakly sensitive to the energy tolerance $\epsilon_E/(J_2/J_1)$. This supports the interpretation that the limiting step is global orientation selection rather than energy relaxation.

Statistical conventions for binomial probabilities, bootstrap boundary estimates, and fitted parameters are summarized in Sec.~\ref{app:uncertainty_estimates}.

\subsection{Dimer winding and sector interpretation}

The dimer-winding diagnostic provides a sector-level interpretation of trapping once \(\rho_{\triangle}^{\rm defect}\ll1\). At finite temperature and for $J_2/J_1>0$, the system is not always exactly inside the nearest-neighbor ground-state manifold. Triangular defects with three equal spins can appear, and when their density is not negligible the dimer winding is no longer a strict topological invariant.

In the low-defect regime, the approach toward stripe order is accompanied by an increase of the stripe-sector proxy $W_{\max}^{\rm proxy}$ toward its maximal value. This is consistent with the idea that the stripe state is not only an ordered configuration but also lies in a sector that is difficult to reach from generic high-temperature states using local updates. The dimer-winding data show that the situation is more subtle than a simple failure to reach the stripe sector. Some failed trajectories already have a large dimer winding, close to a stripe-like sector, while still exhibiting low \(M_{\max}\), large \(\rho_{\rm DW}\), and strong orientation competition. Thus, the limiting step is not only global sector access but also the removal of domain walls and the cleanup of multidomain configurations within or near stripe-like sectors.

The quantitative winding comparison is given in Table~\ref{tab:dimer_winding}. The key point is that failed runs are not necessarily trapped in a central non-stripe sector: some of them already have a large dimer winding but still contain many residual domain walls and competing orientations.

\subsection{Autocorrelation of the dominant stripe order}\label{subsec:autocorrelation_stripe_order}

A fixed-temperature autocorrelation study was performed and summarized in Table~\ref{tab:axe1_autocorr}. Since the autocorrelation function often remains positive up to the maximum lag, these values should be regarded as lower-bound estimates rather than sharply converged equilibrium autocorrelation times.

The magnitude of $\tau_{\mathrm{int}}(M_{\mathrm{stripe}})$ is comparable to $n^*(L,J_2/J_1)$ for $L=128$ and $J_2/J_1=0.09,0.10$. This supports the view that the finite-rate boundary is tied to slow collective memory of the global stripe orientation. It does not invalidate the cooling statistics, because the probabilities reported above are obtained from independent seeds rather than from a single time series assumed to provide independent samples. It does, however, emphasize that the cooling process is strongly out of equilibrium and controlled by collective domain dynamics.

\subsection{Slow-cooling refinement and threshold behavior}\label{subsec:axis3_refine_thresholds}

The refined $L=128$ runs are useful because they test whether the baseline slowest rate, $20000$ sweeps per temperature value, was already sufficient to infer the morphology at weak and intermediate $J_2/J_1$. The corresponding probabilities are tabulated in Appendix~\ref{app:supplementary_tables}, Table~\ref{tab:axis3_refined_probabilities}. At $J_2/J_1=0.05$, the probability remains zero even at $40000$ sweeps per temperature value. This suggests that weak next-nearest-neighbor coupling promotes local stripe organization but does not produce global orientation selection within the present time window.

At $J_2/J_1=0.08$, the single-domain probability increases from $0.12$ at $5000$ sweeps per temperature value to about $0.40$ at $40000$ sweeps per temperature value, but the increase is not sufficient to cross the $0.5$ threshold. The small nonmonotonicity between $20000$ and $40000$ sweeps per temperature value should not be overinterpreted because the trajectory-to-trajectory variance is large and the refined points have finite statistics. The conservative statement is that no systematic conversion to predominantly single-domain stripe order is observed up to $40000$ sweeps per temperature value.

At $J_2/J_1=0.10$, the refined data show a clearer trend. The single-domain probability reaches $0.62$ at $40000$ sweeps per temperature value, so this coupling crosses the operational $P_{\mathrm{single}}\geq0.5$ threshold only at the slowest refined rate. However, it still does not reach $P_{\mathrm{single}}\geq0.9$, so the morphology is not fully saturated at $L=128$.

The resulting threshold summary is given in Appendix~\ref{app:supplementary_tables}, Table~\ref{tab:axis3_refined_thresholds}. The weak-coupling regime $J_2/J_1=0.05$ does not reach even the $P_{\mathrm{single}}\geq0.5$ threshold. The intermediate point $J_2/J_1=0.08$ also remains below this threshold up to $40000$ sweeps per temperature value. The point $J_2/J_1=0.10$ reaches $P_{\mathrm{single}}\geq0.5$ only at $40000$ sweeps per temperature value, while $J_2/J_1=0.20$ reaches the same threshold already at $1000$ sweeps per temperature value and reaches $P_{\mathrm{single}}\geq0.9$ at $20000$ sweeps per temperature value. This hierarchy is consistent with the finite-rate boundary of Sec.~\ref{subsec:finite_rate_stripe_selection}, but it is more restrictive because it requires a real-space single-domain morphology rather than only a global structure-factor threshold.

\subsection{Refined crossover at large system size}\label{subsec:axis4_refined_crossover}

The refined $L=96$ and $L=128$ simulations resolve the transition region between the fully blocked weak-$J_2$ regime and the dynamically accessible stripe regime. Table~\ref{tab:axis4_refined_s5000} shows the refined results at $n_{\mathrm{sweeps}}/T=5000$. The main conclusion is that $J_2/J_1=0.05$ and $0.055$ are still below the practical dynamical threshold at large size. At $L=128$, the strict success probability is only $0.02$ at $J_2/J_1=0.050$ and $0.06$ at $J_2/J_1=0.055$. The onset of substantial stripe success occurs instead near $J_2/J_1=0.065$, where $P_{\mathrm{stripe}}^{E,M}=0.52$ at $L=96$ and $0.54$ at $L=128$.

The weak nonmonotonicity between $J_2/J_1=0.065$ and $0.070$ should not be overinterpreted. With \(50\) seeds, the binomial standard error near \(P=0.5\) is about \(0.07\), corresponding to a \(95\%\) Wilson half-width of about \(0.14\). The refined data are therefore best interpreted as locating a broad finite-time crossover rather than a sharp threshold. Combining the coarse and refined grids gives an operational estimate
\begin{equation}
\left(J_2/J_1\right)^{\star}(n_{\mathrm{sweeps}}/T=5000,L\simeq100\text{--}128) \simeq 0.065\text{--}0.073,
\label{eq:axis4_j2star_estimate}
\end{equation}
where $(J_2/J_1)^\star$ denotes the coupling at which $P_{\mathrm{stripe}}^{E,M}\simeq0.5$ for the present local cooling protocol. This value should not be confused with an equilibrium critical coupling. It is a protocol-dependent dynamical accessibility scale specific to the local Metropolis cooling schedule used here. A nonlocal loop algorithm, an annealing schedule with a longer residence time in the mobility window, or a different microscopic dynamics could shift the numerical value of \((J_2/J_1)^\star\). The robust conclusion is not the precise value of this number but the separation between local constraint restoration, which occurs efficiently, and global stripe-orientation selection, which remains slow under local dynamics.

\begin{figure}[t]
\centering
\includegraphics[width=0.5\textwidth]{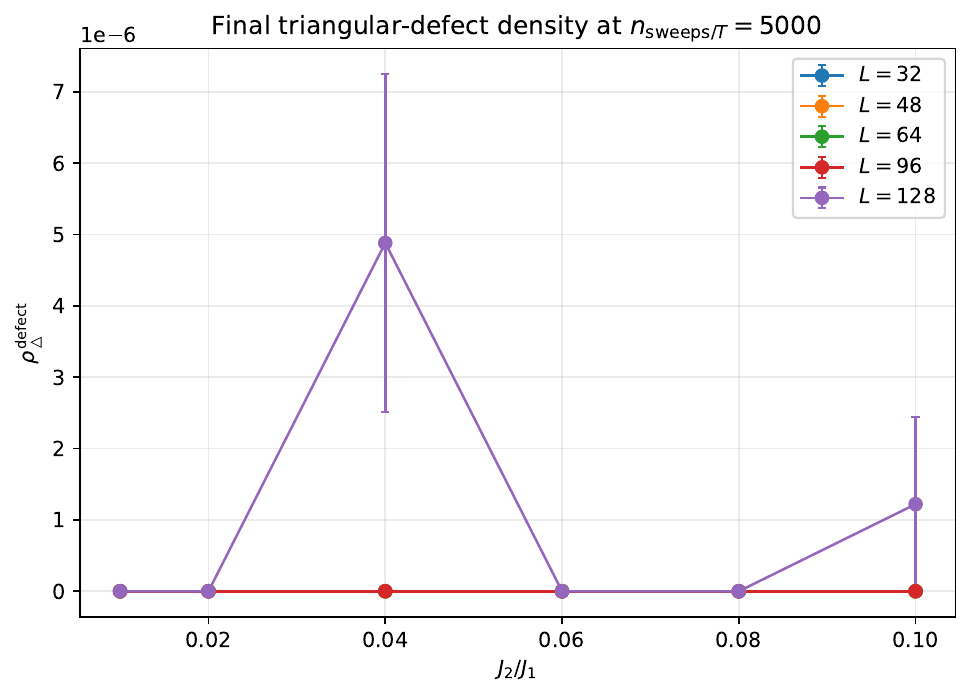}
\caption{Final triangular-defect density at $n_{\mathrm{sweeps}}/T=5000$. The vertical scale is of order $10^{-6}$, showing that local violations of the nearest-neighbor TLIAF constraint are negligible on the scale relevant for stripe selection. The blocked weak-$J_2/J_1$ states are therefore not ordinary thermally excited states with many local constraint violations, but low-defect states close to the frustrated nearest-neighbor manifold.}
\label{fig:axis4_triangle_defects}
\end{figure}

\begin{figure}[t]
\centering
\includegraphics[width=0.5\textwidth]{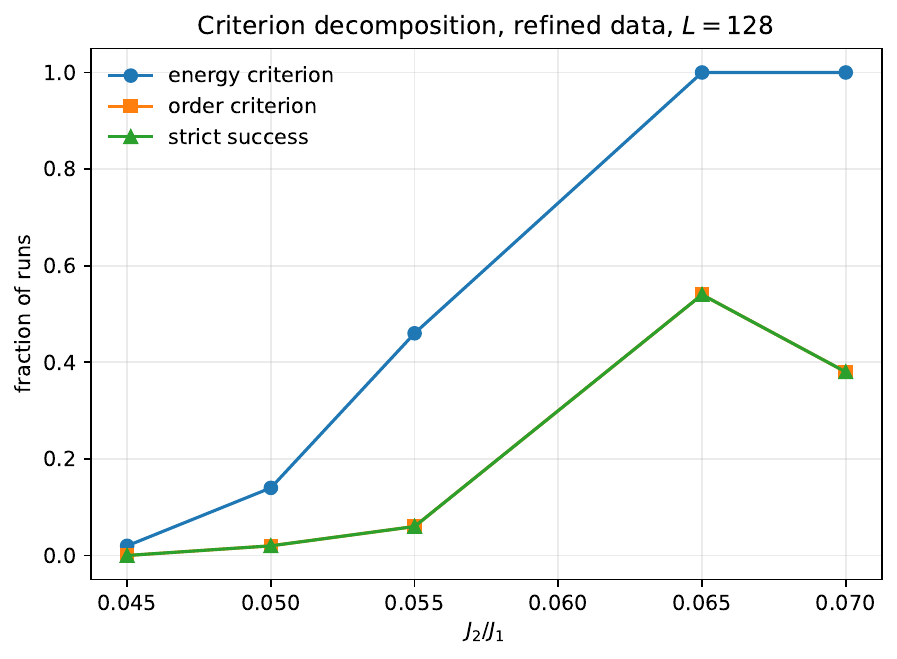}
\caption{Decomposition of the weak-$J_2$ success criterion at $L=128$ and $n_{\mathrm{sweeps}}/T=5000$. The energy criterion is satisfied before the order-parameter criterion in the crossover region. This supports the interpretation that the finite-time bottleneck involves not only energetic relaxation but also the cleanup of orientation defects and multidomain stripe structures.}
\label{fig:axis4_condition_decomposition}
\end{figure}

\section{Supplementary mechanism tables}\label{app:supplementary_tables}

This appendix collects the auxiliary mechanism tables used to support the interpretation of kinetic trapping. These tables document the mechanism run near the kinetic boundary, fixed-temperature autocorrelation estimates, cooling-rate diagnostics, hold protocols, finite-size tests, simple collective-update comparisons, and dimer-winding diagnostics.

\par\medskip
\noindent\begin{minipage}{\columnwidth}
\refstepcounter{table}\label{tab:axe1_boundary_extended}

\centering
\small
\begin{tabular}{c c c c c c c}
\toprule
\(J_2/J_1\) & \(n^*(L=64)\) & \(n^*(L=96)\) & \(n^*(L=128)\) & \(n^*(L=160)\) & \(n^*(L=200)\) & \(n^*(L=256)\) \\
\midrule
0.08 & 3533 & 6119 & 9423 & 27826 & 60000 & -- \\
0.09 & 1331 & 3697 & 6927 & 16700 & 25562 & -- \\
0.10 & 917 & 2470 & 5288 & \(<8000\) & 13775 & 22000 \\
0.11 & 812 & 1500 & 3211 & 5800 & 7750 & -- \\
0.15 & \(<500\) & 630 & 1581 & \(\leq3000\) & 3857 & -- \\
\bottomrule
\end{tabular}

\par\smallskip
\noindent\textbf{Table~\thetable.} {operational stripe-accessibility time \(n^*(L,J_2/J_1)\) obtained from \(P(M_f\geq0.5)=0.5\). Values are expressed in sweeps per temperature value. Inequalities indicate censored estimates, and dashes indicate unsimulated points. The \(L=160,200\) entries at \(J_2/J_1=0.08\) include the additional weak-coupling refinement, while the entries at \(J_2/J_1=0.09\) and \(0.10\) are taken from the refined \(120\)-seed large-size grid. The \(L=256\) entry is a targeted fixed-coupling check at \(J_2/J_1=0.10\).}
\end{minipage}
\par\medskip

\par\medskip
\noindent\begin{minipage}{\columnwidth}
\refstepcounter{table}\label{tab:axis3_refined_probabilities}

\centering
\small
\begin{tabular}{c c c}
\toprule
$J_2/J_1$ & $n_{\mathrm{sweeps}}/T$ & $P_{\mathrm{single}}$ \\
\midrule
0.05 & 20000 & 0.00 \\
0.05 & 40000 & 0.00 \\
0.08 & 5000 & 0.12 \\
0.08 & 10000 & 0.32 \\
0.08 & 20000 & 0.37 \\
0.08 & 40000 & 0.40 \\
0.10 & 5000 & 0.43 \\
0.10 & 10000 & 0.38 \\
0.10 & 20000 & 0.47 \\
0.10 & 40000 & 0.62 \\
\bottomrule
\end{tabular}

\par\smallskip
\noindent\textbf{Table~\thetable.} Combined single-domain probabilities at \(L=128\) after adding the slow-cooling refinements. The entries are central binomial estimates; Wilson confidence intervals are defined in Sec.~\ref{app:uncertainty_estimates}.
\end{minipage}
\par\medskip

\par\medskip
\noindent\begin{minipage}{\columnwidth}
\refstepcounter{table}\label{tab:axis3_refined_thresholds}

\centering
\small
\begin{tabular}{c c c}
\toprule
$J_2/J_1$ & $P_{\mathrm{single}}\geq0.5$ & $P_{\mathrm{single}}\geq0.9$ \\
\midrule
0.05 & not reached & not reached \\
0.08 & not reached & not reached \\
0.10 & 40000 & not reached \\
0.20 & 1000 & 20000 \\
\bottomrule
\end{tabular}

\par\smallskip
\noindent\textbf{Table~\thetable.} Slowest-rate thresholds for single-domain morphology at \(L=128\). The entries give the first simulated value of \(n_{\mathrm{sweeps}}/T\) at which the corresponding probability threshold is reached.
\end{minipage}
\par\medskip

\par\medskip
\noindent\begin{minipage}{\columnwidth}
\refstepcounter{table}\label{tab:axe1_fit_extended}

\centering
\scriptsize
\setlength{\tabcolsep}{2pt}
\renewcommand{\arraystretch}{0.95}
\begin{tabular}{l c c c}
\toprule
Dataset & \(z_{\mathrm{eff}}\) or \(z\) & \(J_c^{\mathrm{kin}}\) & \(\alpha\) \\
\midrule
Global fit, \(L=64,96,128\) & 2.11 & 0.058 & 1.50 \\
Global fit, \(L\leq200\) & \(2.3\text{--}2.5\) & \(\simeq0.06\) & \(\simeq1.5\) \\
Fixed \(J_2/J_1=0.09\) & 2.65 & -- & -- \\
Fixed \(J_2/J_1=0.10,\ L=64\text{--}256\) & 2.30 & -- & -- \\
Fixed \(J_2/J_1=0.10,\ L=128\to256\) & 2.06 & -- & -- \\
\bottomrule
\end{tabular}

\par\smallskip
\noindent\textbf{Table~\thetable.} Phenomenological finite-window estimates for the operational kinetic boundary. The global fits use Eq.~\eqref{eq:kinetic_fit_form} only as empirical interpolations, while the fixed-coupling rows are log--log estimates of \(n^*(L,J_2/J_1)\) versus \(L\). The values are effective parameters for the simulated local Metropolis cooling protocol and should not be interpreted as equilibrium critical exponents or universal dynamical exponents.
\end{minipage}
\par\medskip

\par\medskip
\noindent\begin{minipage}{\columnwidth}
\refstepcounter{table}\label{tab:axe1_mechanism}

\centering
\small
\begin{tabular}{c c c c}
\toprule
\(n_{\mathrm{sweeps}}/T\) & \(\overline{M_f}\) & \(\sigma(M_f)\) & \(P(M_f\geq0.5)\) \\
\midrule
1500 & 0.224 & 0.225 & 0.175 \\
2500 & 0.393 & 0.272 & 0.375 \\
4000 & 0.578 & 0.336 & 0.450 \\
5000 & 0.574 & 0.328 & 0.500 \\
7000 & 0.633 & 0.331 & 0.550 \\
10000 & 0.609 & 0.316 & 0.525 \\
14000 & 0.758 & 0.318 & 0.675 \\
\bottomrule
\end{tabular}
\par\smallskip
\noindent\textbf{Table~\thetable.} Mechanism-run summary for \(L=128\) and \(J_2/J_1=0.10\).
\end{minipage}
\par\medskip

\par\medskip
\noindent\begin{minipage}{\columnwidth}
\refstepcounter{table}\label{tab:L160_L200_J009_J010_refined}

\centering
\small
\begin{tabular}{c c c c}
\toprule
\(L\) & \(J_2/J_1\) & \(n^*\) & Comment \\
\midrule
160 & 0.09 & \(1.67\times10^4\) & interpolated \\
200 & 0.09 & \(2.56\times10^4\) & interpolated \\
160 & 0.10 & \(<8.0\times10^3\) & left-censored \\
200 & 0.10 & \(1.38\times10^4\) & interpolated \\
\bottomrule
\end{tabular}
\par\smallskip
\noindent\textbf{Table~\thetable.} Refined large-size kinetic-boundary estimates for \(L=160,200\) at \(J_2/J_1=0.09,0.10\). Each probability point contains \(120\) independent cooling trajectories. The \(L=160\), \(J_2/J_1=0.10\) entry is left-censored because the first simulated rate already lies above the \(P(M_f\geq0.5)=0.5\) threshold.
\end{minipage}
\par\medskip

\par\medskip
\noindent\begin{minipage}{\columnwidth}
\refstepcounter{table}\label{tab:L256_J010_boundary_summary}

\centering
\small
\begin{tabular}{c c c c c}
\toprule
\(n_{\mathrm{sweeps}}/T\) & \(\overline{M_f}\) & \(\sigma(M_f)\) & \(P(M_f\geq0.5)\) & \(P(M_f\geq0.7)\) \\
\midrule
10000 & 0.390 & 0.313 & 0.32 & 0.21 \\
14000 & 0.420 & 0.290 & 0.30 & 0.18 \\
18000 & 0.506 & 0.313 & 0.48 & 0.29 \\
22000 & 0.539 & 0.344 & 0.50 & 0.34 \\
26000 & 0.567 & 0.349 & 0.51 & 0.36 \\
32000 & 0.612 & 0.346 & 0.57 & 0.42 \\
40000 & 0.614 & 0.339 & 0.53 & 0.44 \\
\bottomrule
\end{tabular}
\par\smallskip
\noindent\textbf{Table~\thetable.} Targeted \(L=256\), \(J_2/J_1=0.10\) finite-rate run. Each point contains \(100\) independent cooling trajectories. The kinetic boundary defined by \(P(M_f\geq0.5)=0.5\) is reached at \(n_{\mathrm{sweeps}}/T\simeq22000\).
\end{minipage}
\par\medskip

\par\medskip
\noindent\begin{minipage}{\columnwidth}
\refstepcounter{table}\label{tab:axe1_autocorr}

\centering
\small
\begin{tabular}{c c c}
\toprule
\(J_2/J_1\) & \(T\) & \(\tau_{\mathrm{int}}\) \\
\midrule
0.09 & 0.10 & 5569 \\
0.09 & 0.12 & 6012 \\
0.10 & 0.10 & 7242 \\
0.10 & 0.12 & 5555 \\
\bottomrule
\end{tabular}
\par\smallskip
\noindent\textbf{Table~\thetable.} Integrated autocorrelation times of \(M_{\mathrm{stripe}}\) in fixed-temperature runs at \(L=128\). Values are in sweeps and should be interpreted as lower-bound estimates when the autocorrelation remains positive up to the maximum lag.
\end{minipage}
\par\medskip

\par\medskip
\noindent\begin{minipage}{\columnwidth}

\refstepcounter{table}\label{tab:axis4_refined_s5000}
\centering
\small
\begin{tabular}{c c c c c c c}
\toprule
$L$ & $J_2/J_1$ & $P$ & $\langle M_f\rangle$ & $\sigma(M_f)$ & $\langle e_f-e_{\mathrm{stripe}}\rangle$ & $(P_E,P_M)$ \\
\midrule
96 & 0.045 & 0.00 & 0.261 & 0.131 & 0.00949 & $(0.00,0.00)$ \\
96 & 0.050 & 0.02 & 0.438 & 0.196 & 0.00781 & $(0.04,0.06)$ \\
96 & 0.055 & 0.10 & 0.549 & 0.203 & 0.00526 & $(0.56,0.14)$ \\
96 & 0.065 & 0.52 & 0.777 & 0.210 & 0.00222 & $(0.98,0.52)$ \\
96 & 0.070 & 0.46 & 0.745 & 0.223 & 0.00222 & $(1.00,0.46)$ \\
128 & 0.045 & 0.00 & 0.216 & 0.128 & 0.00846 & $(0.02,0.00)$ \\
128 & 0.050 & 0.02 & 0.331 & 0.164 & 0.00655 & $(0.14,0.02)$ \\
128 & 0.055 & 0.06 & 0.420 & 0.199 & 0.00552 & $(0.46,0.06)$ \\
128 & 0.065 & 0.54 & 0.746 & 0.246 & 0.00199 & $(1.00,0.54)$ \\
128 & 0.070 & 0.38 & 0.732 & 0.220 & 0.00193 & $(1.00,0.38)$ \\
\bottomrule
\end{tabular}
\par\smallskip
\noindent\textbf{Table~\thetable.} Refined weak-coupling results at $n_{\mathrm{sweeps}}/T=5000$ for $L=96$ and $L=128$. Here $P\equiv P_{\mathrm{stripe}}^{E,M}$, $P_E$ is the energy criterion fraction, and $P_M$ is the order-parameter criterion fraction. The entries are central binomial estimates; Wilson confidence intervals are defined in Sec.~\ref{app:uncertainty_estimates}.
\end{minipage}
\par\medskip

\par\medskip
\noindent\begin{minipage}{\columnwidth}
\refstepcounter{table}\label{tab:axis4_Pstripe_s5000}

\centering
\small
\begin{tabular}{c c c c c c c}
\toprule
$L$ & \multicolumn{6}{c}{$J_2/J_1$} \\
\cmidrule(lr){2-7}
& $0.01$ & $0.02$ & $0.04$ & $0.06$ & $0.08$ & $0.10$ \\
\midrule
32 & 0.00 & 0.00 & 0.00 & 0.84 & 1.00 & 1.00 \\
48 & 0.00 & 0.00 & 0.00 & 0.76 & 0.96 & 1.00 \\
64 & 0.00 & 0.00 & 0.00 & 0.62 & 0.92 & 1.00 \\
96 & 0.00 & 0.00 & 0.00 & 0.40 & 0.66 & 0.98 \\
128 & 0.00 & 0.00 & 0.00 & 0.40 & 0.60 & 0.76 \\
\bottomrule
\end{tabular}
\par\smallskip
\noindent\textbf{Table~\thetable.}Stripe-success probability $P_{\mathrm{stripe}}^{E,M}$ at the slowest coarse cooling rate, $n_{\mathrm{sweeps}}/T=5000$. Each value is obtained from 50 independent seeds. The entries are central binomial estimates, with Wilson intervals defined in Sec.~\ref{app:uncertainty_estimates}.
\end{minipage}
\par\medskip

\par\medskip
\noindent\begin{minipage}{\columnwidth}
\refstepcounter{table}\label{tab:cooling_rate_J208}

\centering
\scriptsize
\setlength{\tabcolsep}{2pt}
\renewcommand{\arraystretch}{0.92}
\begin{tabular}{c c c c c c}
\toprule
\(n_{\mathrm{sweeps}}/T\) & \(P_{\mathrm{stripe}}^{\mathrm{strict}}\) & \(\langle M_{\max}\rangle\) & \(\langle\rho_{\rm DW}\rangle\) & \(\langle C_{\rm orient}\rangle\) & \(\langle W_{\max}^{\rm proxy}\rangle\) \\
\midrule
1000 & 0.50 & 0.705 & 0.195 & 0.250 & 45.13 \\
2000 & 0.68 & 0.808 & 0.128 & 0.191 & 51.70 \\
5000 & 0.67 & 0.850 & 0.100 & 0.150 & 54.40 \\
10000 & 0.89 & 0.946 & 0.036 & 0.054 & 60.56 \\
20000 & 0.96 & 0.975 & 0.016 & 0.025 & 62.42 \\
\bottomrule
\end{tabular}
\par\smallskip
\noindent\textbf{Table~\thetable.} Final-state observables after finite-rate cooling for \(L=64\), \(J_2/J_1=0.08\), and \(100\) independent seeds.
\end{minipage}
\par\medskip

\par\medskip
\noindent\begin{minipage}{\columnwidth}
\refstepcounter{table}\label{tab:Tnuc_J2}

\centering
\small
\begin{tabular}{c c}
\toprule
\(J_2/J_1\) & \(T_{\rm onset}\) \\
\midrule
0.08 & 0.47 \\
0.10 & 0.56 \\
0.15 & 0.74 \\
\bottomrule
\end{tabular}
\par\smallskip
\noindent\textbf{Table~\thetable.} Approximate dynamical ordering-onset temperature extracted from the cooling trajectories at \(L=64\).
\end{minipage}
\par\medskip

\par\medskip
\noindent\begin{minipage}{\columnwidth}
\refstepcounter{table}\label{tab:metastability_J208}

\centering
\scriptsize
\setlength{\tabcolsep}{3pt}
\renewcommand{\arraystretch}{0.92}
\begin{tabular}{c c c c c}
\toprule
\(T\) & Initial state & \(P_{\mathrm{stripe}}^{\mathrm{strict}}\) & \(\langle M_{\max}\rangle\) & \(\langle\rho_{\rm DW}\rangle\) \\
\midrule
0.30 & stripe & 1.00 & 1.000 & 0.000 \\
0.30 & random & 0.44 & 0.699 & 0.198 \\
0.35 & stripe & 1.00 & 1.000 & 0.000 \\
0.35 & random & 0.44 & 0.734 & 0.177 \\
0.40 & stripe & 1.00 & 0.999 & 0.0004 \\
0.40 & random & 0.90 & 0.936 & 0.0425 \\
0.45 & stripe & 1.00 & 0.996 & 0.0030 \\
0.45 & random & 1.00 & 0.996 & 0.0026 \\
\bottomrule
\end{tabular}
\par\smallskip
\noindent\textbf{Table~\thetable.} Fixed-temperature holds for \(L=64\), \(J_2/J_1=0.08\), and \(50\) seeds.
\end{minipage}
\par\medskip

\par\medskip
\noindent\begin{minipage}{\columnwidth}
\refstepcounter{table}\label{tab:size_scaling_trapping}

\centering
\scriptsize
\setlength{\tabcolsep}{3pt}
\renewcommand{\arraystretch}{0.92}
\begin{tabular}{c c c c c}
\toprule
\(L\) & \(P_{\mathrm{stripe}}^{\mathrm{strict}}\) & \(\langle M_{\max}\rangle\) & \(\langle\rho_{\rm DW}\rangle\) & \(\langle C_{\rm orient}\rangle\) \\
\midrule
32 & 0.967 & 0.977 & 0.0155 & 0.023 \\
48 & 0.767 & 0.888 & 0.0748 & 0.110 \\
64 & 0.733 & 0.838 & 0.1078 & 0.161 \\
96 & 0.367 & 0.673 & 0.2102 & 0.311 \\
\bottomrule
\end{tabular}
\par\smallskip
\noindent\textbf{Table~\thetable.} Size dependence of trapping at \(J_2/J_1=0.08\), \(T=0.40\), after \(10^5\) local Metropolis sweeps from random initial conditions.
\end{minipage}
\par\medskip

\par\medskip
\noindent\begin{minipage}{\columnwidth}
\refstepcounter{table}\label{tab:local_vs_nonlocal}

\centering
\scriptsize
\setlength{\tabcolsep}{2pt}
\renewcommand{\arraystretch}{0.92}
\begin{tabular}{c c c c c c c}
\toprule
\(L\) & \(J_2/J_1\) & \(T\) & Update & \(P_{\mathrm{stripe}}^{\mathrm{strict}}\) & \(\langle M_{\max}\rangle\) & \(\langle\rho_{\rm DW}\rangle\) \\
\midrule
64 & 0.10 & 0.35 & local & 0.52 & 0.720 & 0.173 \\
64 & 0.10 & 0.35 & full line & 0.61 & 0.794 & 0.137 \\
64 & 0.10 & 0.35 & segment 64 & 0.61 & 0.779 & 0.145 \\
96 & 0.08 & 0.40 & local & 0.76 & 0.890 & 0.0736 \\
96 & 0.08 & 0.40 & full line & 0.70 & 0.847 & 0.1017 \\
96 & 0.08 & 0.40 & segment 64 & 0.68 & 0.839 & 0.1074 \\
\bottomrule
\end{tabular}
\par\smallskip
\noindent\textbf{Table~\thetable.} Diagnostic comparison between local Metropolis dynamics and simple collective line or segment updates. These collective updates are not optimized loop algorithms and are not used as equilibrium sampling methods.
\end{minipage}
\par\medskip

\par\medskip
\noindent\begin{minipage}{\columnwidth}
\refstepcounter{table}\label{tab:dimer_winding}

\centering
\scriptsize
\setlength{\tabcolsep}{2pt}
\renewcommand{\arraystretch}{0.92}
\begin{tabular}{c c c c c c}
\toprule
Subset & \(\langle M_{\max}\rangle\) & \(\langle\rho_{\rm DW}\rangle\) & \(\langle C_{\rm orient}\rangle\) & \(\langle W_{\max}^{\rm proxy}\rangle\) & \(\langle W_{\max}^{\rm dimer}\rangle\) \\
\midrule
Successful runs & 0.982 & 0.012 & 0.018 & 94.3 & 95.1 \\
Failed runs & 0.598 & 0.268 & 0.401 & 57.4 & 86.4 \\
\bottomrule
\end{tabular}
\par\smallskip
\noindent\textbf{Table~\thetable.} Dimer-winding analysis at \(L=96\), \(J_2/J_1=0.08\), \(T=0.40\), after \(3\times10^5\) sweeps.
\end{minipage}

\clearpage

\par\addvspace{1.5\baselineskip}

\section*{Data and code availability}

The simulation scripts, analysis scripts, processed data, and run seeds used to generate the figures and tables are available from the author upon reasonable request. The programs are in python 3.12. The raw run-level outputs are stored as \texttt{.npz} files, while the aggregated observables used for plotting and statistical analysis are stored as \texttt{.csv} files.

AI was used for this work, helping with learning, programming, writing, correcting, and it has been thoroughly reviewed.

\par\addvspace{1.5\baselineskip}
\bibliographystyle{apsrev4-2}
\bibliography{reference}
\end{document}